\documentclass[11pt]{article}
\usepackage{amssymb,amsmath,amsfonts,mathtools}
\usepackage{enumerate}

\usepackage[sort,compress]{cite}
\usepackage[bulletsep]{collref}
\usepackage{amsfonts,euscript,amssymb,stmaryrd,braket}
\usepackage{graphics,tikz}
\usetikzlibrary{arrows,decorations.markings,patterns}
\usepackage{slashed}
\definecolor{hyperref}{RGB}{026,028,185}
\usepackage[bookmarks=true,colorlinks=true,linkcolor=hyperref,citecolor=hyperref,urlcolor=hyperref,bookmarksnumbered]{hyperref}
\usepackage{cite}

\def\clock{{\count0=\time
           \divide\count0 60
           \ifnum\count0<10 0\fi\the\count0
           \multiply\count0 -60 \advance\count0 \time
           :\ifnum\count0<10 0\fi \the\count0
         }}
\newcommand{\timestamp}{{\small\vbox{\hbox{\tt\jobname.tex}
\hbox{\the\day/\the\month/\the\year, \clock}}}}

\newcommand{\CL}{\mathcal{L}}

\newcommand{\CO}{\mathcal{O}}

\newcommand{\nn}{\nonumber}
\newcommand{\ba}{\begin{eqnarray}}
\newcommand{\ea}{\end{eqnarray}}

\newcommand{\matrto}[4]{\left( \begin{array}{cc} #1 & #2 \\
#3 & #4 \end{array} \right) }

\newcommand{\ads}{\textup{\textrm{AdS}}}

\newcommand{\sphere}{\textup{\textrm{S}}}

\makeatletter
\let\old@startsection=\@startsection
\let\oldl@section=\l@section
\renewcommand{\@startsection}[6]{\old@startsection{#1}{#2}{#3}{#4}{#5}{#6\mathversion{bold}}}
\renewcommand{\l@section}[2]{\oldl@section{\mathversion{bold}#1}{#2}}
\makeatother

\numberwithin{equation}{section}

\def\[{\begin{equation}}
\def\]{\end{equation}}
\newcommand{\be}{\begin{eqnarray}}
\newcommand{\ee}{\end{eqnarray}}

\def\lt{\tilde\lambda}

 \def \t {\tau}
 \def \r {\rho}

\def \O {{\mathcal O}}
\def \m {\mu}

\def \td {\tilde}

\newcommand{\grp}[1]{\mathrm{#1}}

\newcommand{\grSL}{\grp{SL}}

\newcommand{\w}{\mathrm{w}}

\newcommand{\half}{\frac{1}{2}}
\newcommand{\p}{\partial}

\newcommand{\n}{\nu}
\newcommand{\foot}{\footnote}
 
\usepackage{color}

\def \EE {{\mathbb{E}}}
\def \KK {{\mathbb{K}}}

\def\no{\nonumber}
\def\vf{\varphi}

\def\r{\rho}

\def\b{\beta}
\def\k{\kappa}

\DeclareMathOperator{\sn}{sn}

\def\s{\sigma}

\def\r{\rho}
\def\O{\mathcal{O}}

\def\o{\omega}

\def \Om {\Omega }
\def\td{\tilde}
\def \ov {\over}

\def\la{\label}

\begin{document}
\renewcommand{\thefootnote}{\arabic{footnote}}

\overfullrule=0pt
\parskip=2pt
\parindent=12pt
\headheight=0in \headsep=0in \topmargin=0in \oddsidemargin=0in

\vspace{ -3cm} \thispagestyle{empty} \vspace{-1cm}
\begin{flushright} 
\footnotesize
HU-EP-14/29\\
\end{flushright}%

\begin{center}
\vspace{1.2cm}
{\Large\bf \mathversion{bold}
One-loop spectroscopy  
of semiclassically \\ quantized  strings: 
bosonic sector
}

 \vspace{0.8cm} {
  Valentina~Forini$^{a,}$\footnote{ {\tt $\{$valentina.forini,michael.pawellek,edoardo.vescovi$\}$@\,physik.hu-berlin.de}},
 Valentina~Giangreco M. Puletti$^{b,c,}$\footnote{ {\tt vgmp@hi.is}},\\
  Michael~Pawellek$^{a,1}$ and
Edoardo~Vescovi$^{a,1}$}\\
 \vskip  0.5cm

\small
{\em
$^{a}$Institut f\"ur Physik, Humboldt-Universit\"at zu Berlin, IRIS Adlershof, \\Zum Gro\ss en Windkanal 6, 12489 Berlin, Germany  
\vskip 0.05cm
$^{b}$ University of Iceland,
Science Institute,
Dunhaga 3,  107 Reykjavik, Iceland
  \vskip 0.05cm
$^{c}$ Department of Fundamental Physics, Chalmers University of Technology, 412 96 G\"oteborg, Sweden
}
\normalsize


\end{center}

\vspace{0.3cm}
\begin{abstract}

We make a further step in the analytically exact quantization of spinning string states 
in semiclassical approximation, by evaluating the exact one-loop  partition function 
for a class of  two-spin string solutions for which quadratic fluctuations  form a non-trivial system of  
coupled modes. This is the case of a  folded string in the $SU(2)$ sector, 
in the limit  described by a quantum 
Landau-Lifshitz model. The same applies to the full bosonic sector of fluctuations over the folded spinning string in $AdS_5$ with an 
angular momentum $J$ in $S^5$. Fluctuations are
governed by  a special class of fourth-order differential operators,  with  coefficients being meromorphic 
functions on the torus, 
which we are able to solve exactly.

\end{abstract}

\newpage

\tableofcontents

\section{Introduction}

The perturbative approach to string quantization based on semiclassical analysis has proven to be an 
extremely useful tool for investigating the structure of the AdS/CFT correspondence~\cite{Tseytlin:2010jv,McLoughlin:2010jw}.  
Beyond the leading, classical order, direct 2d quantum field theory computations of  string energies are - in general - difficult.
An exception is the case of rational rigid string solutions,  
so-called ``homogeneous''~\cite{Frolov:2003qc,Frolov:2003tu,Frolov:2004bh,Park:2005ji,Beisert:2005mq} 
in that derivatives of the background fields are constant, 
i.e. independent on $(\tau, \sigma)$~\footnote{Non homogenous solutions can become 
homogeneous in certain limits,  as for the  folded string with spin $S$  in $AdS_5$  and  momentum $J$ 
  in $S^5$ in the limit  $\mathcal{S}=\frac{S}{\sqrt{\lambda}}\to\infty$ with $\frac{J}{\sqrt{\lambda}\log S}$
   fixed~\cite{Frolov:2006qe,Roiban:2007jf}. Similarly, in certain cases  one can arrange to make the 
   coefficients in the fluctuation Lagrangian constant~\cite{Hoare:2009rq}.}.  
    In this case the semiclassical analysis is highly simplified since the quadratic fluctuation 
    Lagrangian turns out to have also constant coefficients. Then,  the operator determinants 
    entering the one-loop partition function are expressed in terms of  characteristic frequencies 
    which are relatively simple to calculate,  
    and the computation of 
    quantum corrections can be extended to two-loop order~\footnote{Comments  on higher-loop 
    calculations in such homogenous case are in~\cite{Giombi:2010fa}.} 
    by standard diagrammatic methods~\cite{Giombi:2010fa, Bianchi:2014ada}~\footnote{Another way 
    in which sigma-model perturbation theory has been importantly used in the study of the integrable 
    structure underlying the AdS/CFT system is the calculation of the worldsheet S-matrix, where results exists
    at     tree-level~\cite{Klose:2006zd}, one-loop~\cite{Bianchi:2013nra,Engelund:2013fja,Bianchi:2014rfa,Roiban:2014cia} 
    and two-loop order~\cite{Engelund:2013fja} (for further references see~\cite{Beisert:2010jr}).
}.  

Next to simplest cases are ``non-homogenous'' configurations such as 
rigid spinning string elliptic solutions, the one-spin folded string solution rotating in $\ads_5$~\cite{Gubser:2002tv,Frolov:2002av} 
- being a well-known  example.  This is a
stationary soliton problem for which the classical equations of motion consists in a one-dimensional sinh-Gordon  equation.
In a static gauge where fluctuations along the worldsheet directions are set to zero, fluctuations turn out to 
be governed by differential operators of a  single-gap  Lam\'e type \cite{Beccaria:2010ry}.  Their determinants can be derived explicitly,   
leading to an analytically closed integral expression for the full one-loop string partition function.  
Even if the latter is a complicated integral that is not known how to solve explicitly, a  merit of this analysis 
is to facilitate the investigation of various regimes of interest (BPS or far-from-BPS)
furnishing a ``spectroscopy'' much more precise than the one obtained via a perturbative treatment 
of the fluctuation interactions~\footnote{See for example, in the large spin limit, the detection of 
turning-point contributions for the energy 
of a single-spin string rotating in $AdS_5$~\cite{Beccaria:2010ry} which are missed by naive perturbation theory, 
 or the possibility to check at high orders the peculiar reciprocity-respecting structure of  subleading 
 corrections~\cite{Beccaria:2008tg,Beccaria:2010tb,Forini:2008ky}.}.
This kind of analysis has been then successfully applied also to the single-spin/parameter  case of pulsating string solutions in 
$\ads_5$ and $\mbox{S}^5$~\cite{Beccaria:2010zn}, to  open string duals of space-like Wilson loops describing quark-antiquark 
systems~\cite{Forini:2010ek} or the so-called~\cite{Correa:2012at} Bremsstrahlung function~\cite{Drukker:2011za} 
and to the case of  backgrounds relevant for the $\ads_4$/CFT$_3$ and $\ads_3$/CFT$_2$ correspondence~\cite{Forini:2012bb}.

In the very general case of non-homogenous solutions with more 
than one spin, 
or of single-spin solutions~\cite{Gubser:2002tv,Frolov:2002av} in conformal gauge  
where bosonic fluctuations couple via  the Virasoro constraints~\footnote{In the single-spin case, fermions are  
naturally decoupled at quadratic level. 
In the two-spin case, they couple in a way which mirrors the bosonic sector~\cite{Forini:2012bb}.}, the evaluation of 
the classical energy requires the diagonalization of highly non-trivial second-order \emph{matrix} 2d differential 
operators whose coefficients have a complicated coordinate-dependence. The same 
is true for fluctuations over open string solutions for which the  corresponding  cusped Wilson loops 
have an expectation value which depends on the cusp angle and on another internal 
angle~\cite{Drukker:2011za,Forini:2012bb}. In all these cases the evaluation of the spectrum 
has been performed  setting to zero one of the spins/parameters involved in the 
problem - thus falling back in the category discussed above - or resorting to perturbation theory in them~\cite{Minahan:2005mx}. 
In the case of the single-spin
string, it has been possible to evaluate the exact  one-loop partition function only in static gauge, 
where mixing is absent,  the equivalence with the partition function in conformal gauge being only shown 
numerically~\cite{Beccaria:2010ry}.

Together with the  pedagogical motivation of enriching  the class of problems 
that can be solved analytically, the diagonalization of the mixed-modes 
fluctuation problem for the largest possible set of string configurations is 
interesting for a number of reasons. First, it provides the natural setup for a detailed comparison 
between the algebraic curve approach and the direct worldsheet computation  of energies for string states,
as in some relevant cases passing from one approach to the other implies taking a certain limit which happens to be
non-analytic (see discussion in~\cite{Roiban:2011fe,Gromov:2011de, Fioravanti:2011xw}). Also, it should 
help in the solution of  existing caveats  
for the semiclassical analysis in short string regime~\cite{Roiban:2011fe,Gromov:2011de} as well as in 
the BPS  limit of the ABJM Bremsstrahlung function (see discussion in~\cite{Forini:2012bb, 
Lewkowycz:2013laa, Bianchi:2014laa}).  In general, the classical and the expected quantum 
integrability of the $\ads_5\times \sphere^5$ model should be manifest at the level of small fluctuations 
near a given solution, regardless how complicated is the latter. 
 \bigskip

In this paper we make a first step into the
exact, detailed solution to the  mixed-modes fluctuation spectrum 
in the case of a non-trivial solitonic configuration, the folded string spinning   in $S^5$
with two large angular momenta $(J_1,J_2)$ - as solution of the Landau-Lifshitz (LL)  effective action 
of \cite{Kruczenski:2003gt}~\footnote{See Section \ref{sec:LL} below, for a detailed review 
see \cite{Kruczenski:2004kw,Stefanski:2004cw}.}.  In that this effective model only 
involves a part of the bosonic fluctuations modes (those corresponding to the $SU(2)$ sector), therefore missing
bosonic and fermionic contributions crucial for the UV finiteness of the quantum result for the energy,  it must be 
equipped with an appropriate regularization.   Calculations for the spectrum of the LL model linearized around the 
folded  $SU(2)$ string solution
have been made  in~\cite{Minahan:2005mx} using operator methods via perturbative  evaluation  (in the parameter $J_2/J, ~J=J_1+J_2$) of characteristic 
frequencies, with a sum over them cured via $\zeta$-function regularization. We 
will evaluate here the exact one-loop effective action over the same solution, 
regularized 
by referring the determinants to the limit $J_2=0$, which ensures the expected vanishing of the 
partition function,  and proceeding with a   $\zeta$-function-inspired regularization of the 
path integral.  The ``semiclassically exact analysis'' explained below provides  an efficient 
and elegant tool to find in one step the needed spectral information. 
The result obtained in~\cite{Minahan:2005mx} using perturbation theory up to the $n$-th order 
means here an $n$-th order Taylor expansion. 

The procedure exploits the integrability of a type of fourth-order linear 
differential equations with doubly periodic coefficients~\footnote{A first attempt to study such kind of equations was done by Mittag-Leffler in 
\cite{Mitt}.} in terms of which the bosonic fluctuation 
problem can be usefully re-written, and that emerges as the natural generalization of the Lam\'e differential 
equation. 
Not surprisingly, the same operator appears to govern the mixed-modes bosonic sector of fluctuations 
for the \emph{full} $\ads_5\times \sphere^5$ action when expanded around the folded string solution with 
non-vanishing $\ads_3$ spin and $\sphere^1$ orbital momentum~\cite{Frolov:2002av}. 
As noticed in~\cite{Forini:2012bb}~\footnote{See  Appendix D there.}, the mixing of bosonic fluctuations 
here has its supersymmetric counterpart in a non-trivial fermionic mass matrix.  
Annoyingly,  the differential equations governing the fermionic spectrum  do not satisfy the 
conditions which allowed us to diagonalize the bosonic system, and it is apparently non-trivial 
to find the necessary  generalization of the tools we have developed.  We leave 
the solution of the full string mixed-mode problem for the future~\footnote{Let us underline that the 
quantum Landau-Lifshitz model should, provided an appropriate  regularization prescription is given, 
 still correctly reproduce the full string result~\cite{Beisert:2005mq} while neglecting fermionic fluctuations 
 (as well as a part of the bosonic ones).}, but we 
notice, as nice byproduct of our analysis,  that the tools developed here allow an 
analytic proof of equivalence between the full  (including  fermions~\footnote{This is because 
fermions are decoupled in conformal gauge.}) exact one-loop partition function for the one-spin folded string 
in conformal and static gauge  -- a non-trivial statement which in~\cite{Beccaria:2010ry} has been verified only numerically. 
 \bigskip

The paper proceeds as follows. In Section \ref{sec:generalfolded} 
we present 
the mixed fluctuation problems for the folded string both in the LL and in the full bosonic sectors. 
In Section \ref{sec:4thorder} we discuss 
and construct the solutions of the fourth order differential operator 
governing those spectral problems, which we then solve analytically in  Section \ref{sec:exactpartition}.
Three Appendices follow, collecting details on Sections \ref{sec:generalfolded},  \ref{sec:4thorder}  
and \ref{sec:exactpartition} respectively.

\section{Bosonic fluctuation spectrum for the folded string}
\label{sec:generalfolded}

In this section we consider fluctuations over the classical string configuration 
representing a string solution folded on itself and rotating with two angular 
momenta.
We present the two examples of coupled system of fluctuations 
which we will able to diagonalize in Section \ref{sec:exactpartition}, using the tools presented in Section 
\ref{sec:4thorder}.

\subsection{Landau-Lifshitz fluctuation spectrum for the SU(2) folded string}
\label{sec:LL}
  
The LL effective action, obtained in string theory as a ``fast-string" limit of the Polyakov action
~\cite{Kruczenski:2003gt,Kruczenski:2004cn},
coincides on the gauge theory side with an effective action for the ferromagnetic (bosonic) spin chain in the thermodynamic 
limit~\footnote{The SU(2) sector contains operators of the form $\rm{Tr}(\Phi_1^{J_1}\Phi_2^{J_2})$, 
and the thermodynamic limit reads $J=J_1+J_2\gg 1$.}. 
As such, its role as a bridge between quantum 
string theory and the spin chain description underlying the AdS/CFT Bethe ansatz
has been explored in a number of papers (for a review see \cite{Kruczenski:2004kw,Stefanski:2004cw}).  
We briefly review now the quantum LL approach for the study of a folded 2-spin $(J_1,J_2)$ string 
rotating in $S^5$~\cite{Minahan:2005mx}. 

\bigskip
%
%
The starting point is the LL action for the 
SU(2) sector~\cite{Kruczenski:2003gt,Kruczenski:2004cn}, which is obtained considering a 
string state whose motion with two large spins is restricted to the $S^3$ part of 
$S^5$. 
The collective ``fast'' coordinate ($\beta$)
 associated to the total angular momentum is gauged away, while only transverse ``slow'' coordinates remain to describe the low-energy string motion. 
This is practically implemented by parameterising the 3-sphere coordinates as
$X_1+i X_2=U_1e^{i\beta},~X_3+i X_4=U_2e^{i\beta}\,,~U_aU_a^\star=1$,
fixing the gauge $t=\tau\,,~p_\beta=\text{const}=J$, and rescaling the $t$ coordinate via $\tilde \lambda=\lambda/J^2$, which plays the role of an 
effective parameter. 
%
%
To first order in the $\tilde\lambda$ expansion, 
one gets~\cite{Minahan:2005mx}
\be
\label{initial_L}
&& S_{\rm LL}= J \int d\tau \int_0^{2\pi} {d\s \over 2 \pi} \CL\,, \qquad
\CL=-i U_a^\star \p_\tau U_a-{\lt\over2} \left| D_\s U_a\right|^2 +\CO(\lt^2)\,, ~~\lt\equiv {\lambda \over J^2}\,,\\ \nn
&& D U_a= d U_a-i C U_a \,, \qquad  D U^\star_a= d U^\star_a+ i C U^\star_a \,, \qquad C= -i U_a^\star 
dU_a~.
\ee
%
%
For the two-spin folded string,  describing a closed folded string at the center of AdS, 
at fixed angle in $S^5$, rotating within a $S^3\subset S^5$ with arbitrary frequencies 
$w_1\,, w_2$, the non-vanishing part of the metric is
\[
ds^2=-dt^2+d\psi^2+\cos^2\psi \, d\varphi_1^2+\sin^2\psi \, d\varphi_2^2\,,
\]
and one can write~\cite{Kruczenski:2003gt}
\begin{eqnarray}
ds^2&=& -dt^2+dX_a dX_a^\star\,, \qquad X_a X_a^\star =1\,, \qquad X_a= e^{i \beta} 
U_a\,, \\
U_1&=&\cos\psi \, e^{i\varphi}\,, \qquad U_2= \sin\psi \, e^{-i\vf}\,, \qquad \vf= {\vf_1-\vf_2\over 2}\,, \qquad \beta= {\vf_1+\vf_2\over 2}\,.
\end{eqnarray}
%
%
%
%
%
Hence, the initial Lagrangian \eqref{initial_L} becomes~\cite{Minahan:2005mx}
\[
\label{eq219}
\CL= \cos{2\psi}\,\dot\vf -{\lt\over 2} \left({\psi'}^2+\sin^2{2\psi}{\vf'}^2\right)\,. 
\]


The equations of motion following from \eqref{eq219} are in terms of a 1d sine-Gordon 
equation
\be
&& \psi''+2 \w \sin{2\psi}=0\,, \qquad \vf=-w \,t\,, 
\qquad w={w_2-w_1\over 2}>0\,, \qquad \w={w\over \lt}\,,\\ \nn
&& {\psi'}^2 =2 \w \left(\cos{2\psi}-\cos{2\psi_0}\right)\,,
\ee
whose solution can be written as~\cite{Minahan:2005mx}
\begin{eqnarray}\label{useful_LL}
 \sin\psi(\sigma)&=&k\,\mathrm{sn}(C\sigma,k^2),\qquad \cos\psi(\sigma)=\mathrm{dn}(C\sigma,k^2),\qquad
 k^2=\sin^2\psi_0\,,\\ \nn
  \sqrt{\mathrm{w}}&=&\frac{1}{\pi}\mathbb{K}(k^2),\qquad 
 C=\frac{2}{\pi}\mathbb{K}(k^2)=2\sqrt{\mathrm{w}}~,\qquad\frac{\EE(k^2)}{\KK(k^2)}=1-\frac{J_2}{J}~.
\end{eqnarray}
%
%
The two non-zero spins $(J_1,J_2)$ are
\be
&& J_1
=  w_1 \sqrt{\lambda}\int^{2\pi}_0 {d\s \over 2\pi}\cos^2\psi \,,
\qquad J_2
=w_2 \sqrt{\lambda}\int^{2\pi}_0 {d\s \over 2\pi}\sin^2\psi \,, \qquad
{J_1\over w_1} +{J_2\over w_2} =\sqrt\lambda\,.~~~
\ee


From the Lagrangian \eqref{eq219}, one can expand around the classical solution
\[
\vf= \vf_{cl}+{1\over \sqrt J}\delta\vf(\tau,\sigma)\,, \qquad 
\psi= \psi_{cl}+{1\over \sqrt{J}} \delta\psi(\tau, \sigma)\,,
\]
obtaining, with the field redefinition
\[
f_1 =- \sin(2\psi_{cl}) \delta \vf\,,  ~~~~~~f_2=\delta\psi\,,
\]
and after symmetrization, 
a fluctuation Lagrangian \cite{Kruczenski:2004cn, Frolov:2003qc, Beisert:2003ea} 
which can be usefully written as follows~\cite{Minahan:2005mx}
\be\label{L_LL}
\CL_{LL}=2\,f_2\,\dot{f_1}-\frac{1}{2}\tilde{\lambda}\,\Big[\,f_1'^2+f_2'^2-V_1(\sigma)\,f_1^2-V_2(\sigma) 
\,f_2^2\,\Big]\,.
\ee
Above, in terms of Jacobi functions, one has 
\be\label{def_potentialVi}
 V_1(\s)= 4\w\,\left[1+4 k^2 -6k^2\, \sn^2(C\s,k^2)\right]\,, \qquad 
V_2(\s)= 4\w\, \left[1- 2k^2\, \sn^2(C\s,k^2)\right]\,.~~~
\ee
The time-independence of the potentials allows the Fourier-transform  $\partial_\tau=i\,\omega$, after which the fluctuation equations following from \eqref{L_LL} form the following non-trivial matrix eigenvalue problem  for the characteristic frequencies $\omega$~\footnote{As in \cite{Minahan:2005mx}, the time has been rescaled by $\tilde\lambda$, which we will restore in the final expressions.}
\begin{eqnarray}\label{eq:LLeq1}
 -f_2''(\sigma)-V_2(\sigma)f_2 &=& i\omega f_1\,,\\\label{eq:LLeq2}
  f_1''(\sigma)+V_1(\sigma)f_1 &=& i\omega f_2.
\end{eqnarray}
This system can be solved perturbatively  in the elliptic modulus $k^2$ (or equivalently, in $J_2/J$), 
and so has been done in~\cite{Minahan:2005mx}.  
The main result of this paper is the  \emph{analytically  exact diagonalization} of this non trivial spectral problem. 

\bigskip

To proceed in an analytically exact fashion,
we start by decoupling \eqref{eq:LLeq1}-\eqref{eq:LLeq2} into two fourth-order equations
\begin{eqnarray}\label{eq:4thorder}
 f_2''''+\left[V_1(\sigma)+V_2(\sigma)\right]f_2''+2V_2'(\sigma)f_2'+\left[V_2''(\sigma)+V_1(\sigma)V_2(\sigma)\right]f_2&=&4\omega^2f_2,\\
 f_1''''+\left[V_1(\sigma)+V_2(\sigma)\right]f_1''+2V_1'(\sigma)f_1'+\left[V_1''(\sigma)+V_1(\sigma)V_2(\sigma)\right]f_1&=&4\omega^2f_1,
\end{eqnarray}
which, using operator notation
\begin{equation}\label{def_Oi}
 \mathcal{O}_i=-\frac{\mathrm{d}^2}{\mathrm{d}\sigma^2}-V_i(\sigma),
\end{equation}
 can be compactly written as
\begin{equation}
 \mathcal{O}_1\mathcal{O}_2f_2=\omega^2f_2,\qquad \qquad 
 \mathcal{O}_2\mathcal{O}_1f_1=\omega^2f_1\,.
\end{equation}
The diagonalization of the matrix eigenvalue problem defined by \eqref{eq:LLeq1}-\eqref{eq:LLeq2} is then 
equivalent to the diagonalization of
\be\label{OLL}
\mathcal O_{LL}=  \matrto{\mathcal{O}_1}{~-2\p_\tau}{2\p_\tau~}{~~\mathcal{O}_2}\,,
\ee
where we used the definitions \eqref{def_Oi}. 

Defining  a new coordinate \(x=C\sigma=2\sqrt{\mathrm{w}}\sigma\), 
the first equation of the system \eqref{eq:LLeq1}-\eqref{eq:LLeq2}  can be rewritten as (we define $f_2\equiv f$, and omit in the Jacobi functions the dependence on the modulus $k^2$)
\begin{equation}\label{LL_first}
 \mathcal{O}^{(4)}\,f(x)=0\,,\qquad
\mathcal{O}^{(4)}=\partial_x^{4}+2\,\big(1+2k^2-4k^2\,\mathrm{sn}^2 (x) \big)\,\partial_x^{2}-8k^2\,\mathrm{sn}(x)\,\mathrm{cn}(x)\,
 \mathrm{dn}(x)\,\partial_x+1-\Omega^2\,,
\end{equation}
with
\begin{equation}\label{Omega}
 \Omega=\frac{\omega}{2\mathrm{w}}\equiv\frac{\omega\,\pi^2}{2\,\KK^2}~.
\end{equation}
Equation \eqref{LL_first} is a fourth-order differential equation with doubly-periodic elliptic
coefficient functions~\footnote{For a concise review of the relevant properties and identities for Jacobi elliptic functions  see for example Appendix A of \cite{Beccaria:2010ry}.}
with period $2L=4\KK$ (following from the $2\pi$-periodicity of the closed string) and only one regular singular pole, in Fuchsian classification.  
A first (incomplete) attempt to study this kind of equations was done by 
Mittag-Leffler~\footnote{Or by the student he mentions in a footnote of~\cite{Mitt}.} in 
\cite{Mitt}, and to our knowledge not much else is known in literature. 
In Section  \ref{sec:4thorder} we will present a systematic study of the eigenvalue problem associated 
to this equation, showing that the corresponding determinant can be computed 
analytically.
Before doing that we show that this specific class of operators is of more 
general interest, as it appears governing (at least in the bosonic case) the 
 spectrum of fluctuations
above the folded string with two angular momenta~\cite{Frolov:2002av}, and thus it can likely be of help for the  study of a large variety
of  problems involving a  
coupled system of fluctuations above elliptic string solutions~\footnote{This observation is based on the 
already noticed similarity between  the fluctuation spectra over the minimal surfaces corresponding space-like Wilson 
loops of~\cite{Drukker:2011za} and the  one of ~\cite{Frolov:2002av,Beccaria:2010ry}.}.

 \subsection{Folded string in full bosonic sigma-model}

Quadratic fluctuations over a folded   string solution rotating with two angular momenta $(S,J)$  in 
$\ads_5$ and in $\sphere^5$~\cite{Frolov:2002av} are non-trivially coupled both in their bosonic sector~\cite{Frolov:2002av} and 
in the fermionic one~\cite{Forini:2012bb}, and regardless of the gauge 
choice.  Notice the different conventions to label frequencies and parameters characterising the classical solution between our work and \cite{Frolov:2002av, Beccaria:2010ry}. 

\subsubsection{Bosonic sector}
\label{sec:foldedfullbos}

Bosonic fluctuations over the classical closed string solution
\begin{eqnarray}\label{solcl1}
t&=&\kappa\,\tau\,,\qquad\qquad
\phi=\bar w\,\tau\,\qquad\qquad\varphi=\nu\,\tau\,,\qquad\qquad\kappa, \bar w,\nu={\rm const}\,,\\ 
\label{solcl2}
\rho&=& 
\rho(\sigma)=\rho(\sigma+2\pi)\,,\qquad \beta_u=0\,,\,(u=1,2)\,,\qquad 
\psi_s=0\,,\,(s=1,2,3,4)\,,
\end{eqnarray}
with $(t,\rho,\phi,\beta_u)$ describing  $\ads_5$ and $(\varphi,\psi_s)$ spanning $S^5$, are described 
in conformal gauge by the following 
Lagrangian~\cite{Frolov:2002av}
\begin{eqnarray}\no
 {\cal L}_B^{\rm folded}&=&- \partial_a \td {t} \partial^a \td {t}- \mu_t^2 \td {t}^2 +   \partial_a \td {\phi}
 \partial^a \td {\phi}+ \mu_{\phi}^2 \td {\phi}^2 +\partial_a \td {\rho} \partial^a \td {\rho}+\mu_{\rho}^2 \td {\rho}^2+ 4\, \td {\rho} (\kappa \sinh \rho\ \partial_0 \td {t} - \bar w\,\cosh \rho\ \partial_0 \td {\phi})
  \\
&&+ \, \partial_a {\tilde\beta}_u \partial^a {\tilde\beta}_u +\mu_{\beta}^2 {\tilde\beta}_u^2+\p_a\td{\varphi}\p^a\td{\varphi}
+\p_a\td{\psi}_s\p^a\td{\psi}_s+\nu^2\,\td{\psi}^2_s\ .   
 \label{lagconf}
 \ea
Here the fields with tildes are fluctuations over the background 
\eqref{solcl1}-\eqref{solcl2}, and have 
masses
\ba 
\mu_t^2&= 2 \rho'^2 -\kappa^2+\nu^2, ~~~~~~~~~~~~~~~~~~~~~~~~~~ \mu^2_{\phi}&=2 \rho'^2 -\bar{w}^2+\nu^2, \\\nn
\quad \mu^2_{\rho}&=2 \rho'^2 - \bar{w}^2-\kappa^2+2\nu^2,~~~~~~~~~~~~~~
\quad \mu_{\beta}^2&=2 \rho'^2+\nu^2  
\label{massesconf}
\end{eqnarray}
given in terms of  the non-trivial classical field $\rho$ satisfying the equation of motion 
($\KK\equiv \KK (k^2)$)
\ba\label{eom}
\rho'^2 &=& \kappa^2\cosh^2\rho-\bar{w}^2\sinh^2\rho-\nu^2\ , \ \ \ \ \ \ \ \ 
k^2=\frac{\kappa^2-\nu^2}{\bar{w}^2-\nu^2}
\\\label{rhop}
\rho'^2 (\sigma) &=&  (\kappa^2-\nu^2)\,{\rm sn}^2(\sqrt{\bar{w}^2-\nu^2}\,\sigma+\KK\,|\,k^2)\ .
\ea
 The $\tilde\beta_u$ fluctuating fields, transverse to the motion of the  
classical solution and decoupled from the other but with nontrivial mass, give a contribution to the 
one-loop partition function that has been evaluated exactly in~\cite{Beccaria:2010ry}. 

The remaining three $\ads_3$ fields $(t,\rho,\phi)$ and the $\varphi$ field 
in $S^5$ are non-trivially coupled through Virasoro constraints. Their equations of motion 
read   
\ba\label{eom1}
&&(\p^2_\t-\p^2_\s)\,\td t+\mu^2_{ t}\,\td t+2\,\k\,\sinh\r\,\p_\t\td \r=0\,,\\\label{eom2}
&&(\p^2_\t-\p^2_\s)\,\td\r+\mu^2_{ \r}\,\td\r+2\,(\k\,\sinh\r\,\p_\t\td  t-\bar w\,\cosh\r\,\p_\t\td
 \phi)=0\,,\\\label{eom3}
&&(\p^2_\t-\p^2_\s)\,\td \phi+\mu^2_{\phi}\,\td  \phi+2\,\bar w\,\cosh\r\,\p_\t\td\r=0 \ , 
\ea
together with the free field equation for $\td\varphi$.
From the  conformal gauge conditions (Virasoro constraints) it follows
\ba\label{v}
&&-\k\,\cosh^2\r\,\p_\t\,{\td  t} +  (\bar{w}^2-\k^2)\,\sinh\r\,\cosh\r\,\td\r+
\nu\,\p_\tau\td\varphi+\r'\,\p_\s\,\td\r+\bar w\,\sinh^2\r\,\p_\t\,{\td
 \phi}=0\,, ~~~~\\\label{vi}
&&-\k\,\cosh^2\r\,\p_\s\,{\td t}+\bar w\,\sinh^2\r\,\p_\s{\td \phi}+\nu\,\p_\sigma\td\varphi+\r'\,\p_\t\,\td\r=0 \ . 
\ea
Since the $\r$-background does not depend on $\t$ and since the above equations are linear 
we may consider to pass at the Fourier  mode level, i.e. replacing $\td t\to e^{i\,\omega\,\tau}\td t$ 
e $\phi\to e^{i\,\omega\,\tau}\td\phi$. 
Then the Virasoro constraints imply  for $\td t$ and $\td \phi$ (not yet switching to Euclidean)
\ba\label{tfromVIR}
\td t&=& \frac{\nu\,\cosh\r}{\k}\,\td\varphi+
\frac{i\,\sinh\r}{2\,\k\,\omega}\,\Big[\p^2_\s-2\r'\,\coth\r\,\p_\s+(\omega^2+\k^2-\bar{w}^2)\Big]\,\td\r\,, \\\label{phifromVIR}
\td\phi&=&\frac{\nu\,\sinh\r}{\bar w}\,\td\varphi+
\frac{i\,\cosh\r}{2\,\bar w\,\omega}\,\Big[\p^2_\s-2\r'\,\tanh\r\,\p_\s+(\omega^2-\k^2+\bar{w}^2)\Big]\,\td\r\,.
\ea
Substituting in \eqref{eom1}-\eqref{eom3} the expressions (\ref{tfromVIR}) and (\ref{phifromVIR}) we get
that one of them is satisfied automatically while the other two become equivalent 
(they differ only up to the free equation of motion for $\tilde\varphi$) to the following equation
\be\label{eq1}
{\cal O}^{(4)}  \td \r=-4\,i\, \n \, \omega \, \r'\, \p_\s\td\varphi
\ee
where
\be \label{Ofourthnu}
{\cal O}^{(4)} = \frac{1}{\r'}(\partial^2_\s+\o^2-V(\sigma))\,\r'^2\,(\partial^2_\s
+\o^2)\frac{1}{\r'}-4\,\nu^2\,\o^2 
\ee
with 
\be
V(\sigma)=2\,\rho'^2+2\frac{(\k^2-\nu^2)(\bar w^2-\nu^2)}{\r'^2}~.
\ee
Being $\td\varphi$ a free field one can write (\ref{eq1}) as~\footnote{This structure has been understood in 
collaboration with M. ~Beccaria, G.~ Dunne and A.~A.~Tseytlin.}
\be\label{eq2}
(\partial^2_\sigma+\o^2)\,\frac{1}{\rho'}\,{\cal O}^{(4)}  \td \r=0.
\ee

Changing to Euclidean signature, \(\o^2\to-\o^2\) and introducing the new coordinate
\(x=\sqrt{\bar{w}^2-\nu^2}\sigma\) the operator 
gets the canonical form
\begin{equation}\label{eq:opnunonzero2}
 \mathcal{O}^{(4)}=\partial_x^4+2[-\bar{\Omega}^2+k^2+1-4k^2\mathrm{sn}^2(x)]\partial_x^2-8k^2\mathrm{sn}(x)\mathrm{cn}(x)\mathrm{dn}(x)\partial_x
 +[(\bar{\Omega}^2+1+k^2)^2-4k^2]+\frac{4\nu^2\bar{\Omega}^2}{\bar{w}^2-\nu^2},
\end{equation}
where  we used the short notation 
\be\label{baromega} 
\bar{\Omega}^2=\frac{\o^2}{\bar{w}^2-\nu^2}.
\ee 
The operator above is strikingly similar to the one \eqref{LL_first} emerging in the LL quantum model, 
 displaying however a significative difference as it cannot be seen as  a ``traditional''
eigenvalue problem. Indeed,  $\bar\Omega$ does not only  appear in the constant term but also in the coefficient of the  second-order derivative  
\footnote{Such case is apparently called in literature as ``polynomial operator 
pencil''~\cite{FursevV}.}. In Appendix \ref{app:foldedfullstatic} we show that the same operator appears in static gauge.

It should be noticed that even in the simpler  case of single spin, with  the folded string solution only rotating
in $\ads$, bosonic fluctuations are still coupled in conformal gauge with a fourth order operator which 
can be easily obtained from \eqref{eq:opnunonzero2} setting $\nu=0$.

\subsubsection{Fermionic sector}
\label{sec:foldedfermions}

As found in~\cite{Forini:2012bb} (see Appendix D there),  in the two-spins case to the  coupled system of bosonic
modes  corresponds a non trivial fermionic mass matrix. 
After  two local boosts and a standard field redefinition~\footnote{See Appendix D in~\cite{Forini:2012bb}.} the fermionic fluctuation Lagrangian can be put in the form $
{\cal L}_F=2\,\bar\psi\,D_F\,\psi~,
$
where
\be\label{fermop}
 D_F=i\,\Big[\,\Gamma^a\,\partial_a+a(\sigma)\,\Gamma_{234}+b(\sigma)\,\Gamma_{129}~\Big],
\ee
with
\be\label{ab}
a(\sigma)=-\sqrt{\rho'^2+\nu^2}, \quad\quad\quad b(\sigma)=\frac{\nu\,\kappa\,\bar w}{2(\rho'^2+\nu^2)}~.
\ee
Squaring (\ref{fermop}), one obtains
\be\label{DF2}
\!\!\!\!\!\!\!\!D_F^2=-\partial_a\partial^a-\Gamma_{29}\,\Big(2\,b(\sigma)\,\partial_\sigma+b'(\sigma)\Big)+a^2(\sigma)+b^2(\sigma)-a'(\sigma)\,\Gamma_{1234}+2a(\sigma)\,b(\sigma)\,\Gamma_{1349}~.
\ee
Noticing that the matrices appearing in (\ref{DF2}) satisfy, together with their product, a 2-dimensional Dirac algebra,
\be
\!\!\!\!\!\!\!\!\{\Gamma_{29}, \Gamma_{1234}\}=0=\{\Gamma_{29}, \Gamma_{1349}\}=\{\Gamma_{1234}, \Gamma_{1349}\}~, \quad\quad\quad
\Gamma_{1234}^2=\Gamma_{1349}^2=-\Gamma_{29}^2=\mathbf{1_{32}}~,
\ee
one may therefore choose a representation in which 
\be
\Gamma_{29}=i\,\sigma_2 \times \mathbf{1_8},~~~~~\Gamma_{1234}=\sigma_3\times  \mathbf{1_8},~~~~~ \Gamma_{1349}=\sigma_1\times\mathbf{1_8}~,
\ee  
where $\sigma_i$ are Pauli matrices. The fermionic fluctuations are then equivalent to 8 copies of coupled fields $\psi_1,\psi_2$ whose equations of motion, going to Euclidean space and Fourier transforming in $\tau$, read
\ba\label{eqferm1}
&&\Big(-\partial^2_\sigma+\omega^2+a^2+b^2-a'\Big)\,\psi_1-\Big(2b\,\partial_\sigma+b'-2a\,b\Big)\,\psi_2=0\\\label{eqferm2}
&&\Big(-\partial^2_\sigma+\omega^2+a^2+b^2+a'\Big)\,\psi_2+\Big(2b\,\partial_\sigma+b'+2a\,b\Big)\,\psi_1=0~.
\ea
It is possible to decouple the above equations and write a fourth-order differential equation for example for $\psi_1$.
However, we do not report here its lengthy expression, as its coefficient functions are not
meromorphic functions on the torus, which is the kind of differential operator we 
have been able to solve with the method exposed in Section \ref{sec:4thorder}. 
In particular, it is the presence of the function $a(\sigma)=-\sqrt{\rho'^2+\nu^2}$ which introduces 
branch-cuts   ruining  the simple pole structure  at the basis of the procedure described below in Section 
\ref{polestructure}.  The study of a suitable generalization of such procedure does not appear to be 
trivial~\footnote{Also, the use of local  target space rotations  of the type already used, 
for example, in~\cite{Frolov:2002av} or~\cite{Drukker:2011za,Forini:2012bb} 
does not lead to any simplification  of the  coefficients.}.

 We conclude this section recalling that in the single-spin case ($\nu=0$, therefore $b=0$) 
the system \eqref{eqferm1}-\eqref{eqferm2}
decouples  giving two second-order Lam\'e type equations which differ only due to the $\pm a'$ term. 
The corresponding fermionic functional determinants (which in fact coincide)  
have been evaluated exactly in~\cite{Beccaria:2010ry}.

\section{Fourth order linear differential equations with doubly periodic coefficients}
\label{sec:4thorder}

In this Section we study the properties  of fourth order differential equations with doubly periodic coefficients. 
We first generalize the Floquet analysis of second order linear differential equations with 
periodic coefficients as done in \cite{Magn}, then find  the explicit solution for the specific  class of operators 
of interest in this paper. 
Because of the striking similarity of the Bloch solutions \eqref{eq:ansatz} and quasi-momentum finite-gap 
structure \eqref{finitegap} found here with the corresponding ones in the second-order decoupled 
case studied in~\cite{Beccaria:2010ry} (see also \cite{Beccaria:2010zn}), we can define the operators here 
analyzed as a higher-order generalization of the second order finite-gap Lam\'e equation.

\subsection{Floquet theory of determinants of fourth order one-dimensional operators}

Consider the fourth order differential operator 
\begin{equation}
 \mathcal{O}^{(4)}=\partial_x^4+v_1(x)\partial_x^2+v_2(x)\partial_x+v_3(x),
\end{equation}
where the coefficient functions \(v_i(x)\) have a fundamental period \(L\), 
\begin{equation}
 v_i(x+L)=v_i(x)~.
\end{equation}
The solution of the corresponding eigenvalue problem
\begin{equation}\label{eq:diffeq}
 \mathcal{O}^{(4)}f(x)=\Lambda f(x),\qquad f(x+L)=f(x)~
\end{equation}
consists of four independent functions \(f_i(x)\), which can be
normalized as
\begin{eqnarray}
 f_1(0)=1,\qquad f_1'(0)=0,\qquad f_1''(0)=0,\qquad f_1'''(0)=0,\\\nn
 f_2(0)=0,\qquad f_2'(0)=1,\qquad f_2''(0)=0,\qquad f_2'''(0)=0,\\\nn
 f_3(0)=0,\qquad f_3'(0)=0,\qquad f_3''(0)=1,\qquad f_3'''(0)=0,\\\nn
 f_4(0)=0,\qquad f_4'(0)=0,\qquad f_4''(0)=0,\qquad f_4'''(0)=1~.
\end{eqnarray}
The Wronskian determinant is therefore normalized to 
\begin{equation}\label{eq:Wronskian}
 W=f_1(0)f_2'(0)f_3''(0)f_4'''(0)=1~.
\end{equation}
Given a complete set of solutions \(f_i(x)\),  the periodicity of (\ref{eq:diffeq}) implies that also \(f_i(x+L)\) is a solution,
which can be written as a linear combination of the \(f_i(x)\):
\begin{equation}
 f_i(x+L)=\sum_{j=1}^4a_{ij}f_j(x),\qquad i=1,2,3,4\,.
\end{equation}
Setting \(x=0\) one gets
\be
a_{ij}= f^{(j-1)}_i (L)\,,
\ee
with $f^{(0)}_i=f_i$\,, 
which defines the monodromy matrix
\begin{equation}
 \mathcal{M}(\Lambda)=\left(\begin{array}{cccc} f_1(L) & f_1'(L) & f_1''(L) & f_1'''(L) \\ \\ f_2(L) & f_2'(L) & f_2''(L) & f_2'''(L) \\ \\
 f_3(L) & f_3'(L) & f_3''(L) & f_3'''(L) \\  \\ f_4(L) & f_4'(L) & f_4''(L) & f_4'''(L)\end{array}\right)\,.
\end{equation}
By diagonalizing this matrix one obtains a new set of four linear independent 
solutions \(\bar f_i(x)\), the Floquet or Bloch solutions with the
property \(f_i(x+L)=\rho_if_i(x)\), or
\begin{equation}
 \left(\begin{array}{cccc} f_1(L)-\rho & f_1'(L) & f_1''(L) & f_1'''(L) \\ \\ f_2(L) & f_2'(L)-\rho & f_2''(L) & f_2'''(L) \\ \\
 f_3(L) & f_3'(L) & f_3''(L)-\rho & f_3'''(L) \\  \\ f_4(L) & f_4'(L) & f_4''(L) & f_4'''(L)-\rho\end{array}\right)\left(\begin{array}{c}
 \bar f_1(x) \\  \\ \bar f_2(x) \\  \\ \bar f_3(x) \\ \\ \bar f_4(x)\end{array}\right)=0.
\end{equation}
As usual, this equation has non-trivial solutions \(\bar f_i(x)\) provided  that \(\det(\mathcal{M}-\rho\mathbf{1})=0\), with 
\begin{eqnarray}\nonumber
 &&\det(\mathcal{M}-\rho\mathbf{1})=\rho^4-(f_1(L)+f_2'(L)+f_3''(L)+f_4'''(L))\rho^3+\Big[\det\Big(\begin{array}{cc} f_1 & f_1'\\f_2 & f_2'
 \end{array}\Big)+\det\Big(\begin{array}{cc} f_1 & f_1''\\f_3 & f_3''\end{array}\Big)+\\\label{eq:det0}
 &&+\det\Big(\begin{array}{cc} f_1 & f_1'''\\f_4 & f_4'''\end{array}\Big)+
 \det\Big(\begin{array}{cc} f_2' & f_2''\\f_3' & f_3''\end{array}\Big)+\det\Big(\begin{array}{cc} f_2' & f_2'''\\f_4' & f_4'''\end{array}\Big)+
 \det\Big(\begin{array}{cc} f_3'' & f_3'''\\f_4'' & f_4'''\end{array}\Big)\Big]\rho^2+\\\nonumber
&&-\left[\det\left(\begin{array}{ccc} f_1 & f_1' & f_1''\\f_2 & f_2' & f_2'' \\ f_3 & f_3' & f_3'''\end{array}\right)+
 \det\left(\begin{array}{ccc} f_1 & f_1' & f_1''\\f_2 & f_2' & f_2''' \\ f_4 & f_4' & f_4'''\end{array}\right)+  \det\left(\begin{array}{ccc} f_1 & f_1'' & f_1'''\\f_3 & f_3'' & f_3''' \\ f_4 & f_4'' & f_4'''\end{array}\right)+
 \det\left(\begin{array}{ccc} f_2' & f_2'' & f_2'''\\f_3' & f_3'' & f_3''' \\ f_4' & f_4'' & f_4'''\end{array}\right)\right]\rho
+1,
\end{eqnarray}
where we have used (\ref{eq:Wronskian}).  
Let \(\rho_i\), \(i=1,2,3,4\) be the roots of this fourth order polynomial equation. Then, we can write
\begin{eqnarray}\label{eq:det1}
 \det(\mathcal{M}(\Lambda)-\rho\mathbf{1})&=&\rho^4-(\rho_1+\rho_2+\rho_3+\rho_4)\rho^3+(\rho_1\rho_2+\rho_1\rho_3+\rho_1\rho_4+
 \rho_2\rho_3+\rho_2\rho_4+\rho_3\rho_4)\rho^2-\nonumber\\
 & &-(\rho_1\rho_2\rho_3+\rho_1\rho_2\rho_4+\rho_1\rho_3\rho_4+\rho_2\rho_3\rho_4)\rho+\rho_1\rho_2\rho_3\rho_4.
\end{eqnarray}
Comparing the expressions (\ref{eq:det0}) and (\ref{eq:det1}) gives a condition on the Floquet factors
\begin{equation}\label{eq:conditionFloquet}
 \rho_1\rho_2\rho_3\rho_4=1.
\end{equation}
A general solution to the equation above would require the introduction
of three functions. We can however proceed conveniently introducing just \emph{two} quasi-momenta functions \(p_i(\Lambda)\), \(i=1,2\) with
\begin{eqnarray}\label{eq:Floquetfactors}
 \rho_1&=&e^{ip_1(\Lambda)L},\qquad \rho_2=e^{-ip_1(\Lambda)L}, \nonumber\\
 \rho_3&=&e^{ip_2(\Lambda)L},\qquad \rho_4=e^{-ip_2(\Lambda)L}\,.
\end{eqnarray}
Knowing the quasi-momenta allows us to immediately compute the determinants. In particular, we have
\begin{enumerate}[i)]
\item \underline{Functions with period \(L\):}

Periodic eigenfunctions \(f_i(x+L)=f_i(x)\) exist only for special values of \(\Lambda\) which are determined by
setting \(\rho=1\) in (\ref{eq:det1}) and using (\ref{eq:Floquetfactors})
\begin{eqnarray}
 \mathrm{det}_{P,L}\mathcal{O}^{(4)}&=&
 4-4\cos(p_1L)-4\cos(p_2L)+4\cos(p_1L)\cos(p_2L)=\nonumber\\
 &=&16\sin^2\left(\frac{L}{2}p_1(\Lambda)\right)\sin^2\left(\frac{L}{2}p_2(\Lambda)\right)\,.
\end{eqnarray}
\item
\underline{Anti-periodic functions by \(L\):}

We get the determinant for antiperiodic eigenfunctions \(f_i(x+L)=-f_i(x)\) by setting \(\rho=-1\) in (\ref{eq:det1}) and using (\ref{eq:Floquetfactors})
\begin{eqnarray}
 \mathrm{det}_{AP,L}\mathcal{O}^{(4)}&=&4+4\cos(p_1L)+4\cos(p_2L)+4\cos(p_1L)\cos(p_2L)=\nonumber\\
 &=&16\cos^2\left(\frac{L}{2}p_1(\Lambda)\right)\cos^2\left(\frac{L}{2}p_2(\Lambda)\right)\,.
\end{eqnarray}
\item\underline{Functions with period \(2L\):}

In this case one has to take the product of the previous two determinants, which gives
\begin{equation}\label{detop}
 \mathrm{det}_{P,2L}\mathcal{O}^{(4)}=\mathrm{det}_{P}\mathcal{O}^{(4)}\mathrm{det}_{AP}\mathcal{O}^{(4)}=16\sin^2\left(Lp_1(\Lambda)\right)\sin^2\left(Lp_2(\Lambda)\right)\,.
\end{equation}
 \end{enumerate}

\subsection{Construction of the solutions: a Hermite-Bethe ansatz}
\label{Mittag}
In this section we find the Bloch solutions for a certain class of fourth order periodic differential equations, 
which  we will argue in \ref{app:spectral} to be 
higher order generalizations of the second order finite-gap Lam\'e 
equation. A first attempt to study this kind of equations was done by 
Mittag-Leffler in~\cite{Mitt}. 

In the following we will use the fact that an elliptic function without any poles in a fundamental period parallelogram of 
the complex plane is merely a constant \cite{Whit}.  The differential operators of interest are of the type
\begin{eqnarray}\label{Oholy}
 \mathcal{O}&=&\frac{\mathrm{d}^4f}{\mathrm{d}x^4}+v_1(x)\frac{\mathrm{d}^2f}{\mathrm{d}x^2}+v_2(x)\frac{\mathrm{d}f}{\mathrm{d}x}+
 v_3(x)f(x),
\end{eqnarray}
where the ``potentials''
\begin{eqnarray}\label{potentials}
 v_1(x)&=&\alpha_0+\alpha_1k^2\mathrm{sn}^2(x),\\\nn
 v_2(x)&=&\beta_0+\beta_1k^2\mathrm{sn}^2(x)+2\beta_2k^2\mathrm{sn}(x)\mathrm{cn}(x)\mathrm{dn}(x),\\ \nonumber
 v_3(x)&=&\gamma_0+2\gamma_3k^2+(\gamma_1-4(1+k^2)\gamma_3)k^2\mathrm{sn}^2(x)+2\gamma_2k^2\mathrm{sn}(x)\mathrm{cn}(x)\mathrm{dn}(x)
 +6\gamma_3k^4\mathrm{sn}^4(x),
\end{eqnarray}
are given by elliptic functions \emph{with only one regular singular pole} (in Fuchsian classification) 
at \(x=i\mathbb{K}'\).
The coefficients \(\alpha_0,\alpha_1,\beta_0,\beta_1,\beta_2,\gamma_0,\gamma_1,\gamma_2,\gamma_3\) are so far free parameters.
We will now find conditions on these parameters such that the following eigenvalue equation
\begin{equation}
  \mathcal{O}f(x)=\Lambda f(x), \qquad f(x+L)=f(x)
\end{equation}
is solved by a Hermite-Bethe-like ansatz~\cite{Whit}~\footnote{
The ansatz \eqref{eq:ansatz} provides four linearly independent solutions - see for example \eqref{eq:foursolutionsfolded2}-\eqref{alphafoldedfin}, or Fig. \ref{fig:fundomain} which gives a graphical representation of them.  However, at the edges (a finite set of points) where the color lines meet, there can be a problem, since two or all four functions become linearly dependent. 
This is expected from the second-order case~\cite{Whit,Magn}, where the ansatz gives all two linear
independent solutions, except for a finite number of problematic points (the 'band edge solutions' for 
Lam\'e operators~\cite{Beccaria:2010ry}). The  missing solutions at those points can be found, see the procedure in~\cite{Ince}, and this is expected to be generalizable to our fourth-order case. For our purpose of evaluating a partition function, it is sufficient - see discussion below \eqref{Gamma} - the knowledge of the solutions (in terms of associated quasi-momenta) in the physical region $\Omega^2<0$ which is free - see \eqref{alphabranchnegag} - from such problematic "edge points". 
}
\begin{equation}\label{eq:ansatz}
 f(x)=\prod_{r=1}^n\frac{H(x+\bar\alpha_r)}{\Theta(x)}e^{x\rho}e^{x\lambda}~\,.
\end{equation}
The constants $\rho$ and $\bar\alpha_r$ are determined by analyticity constraints on the eigenfunction as follows. Let us introduce the function $F$
\begin{equation}\label{F}
 F(x)= \frac{1}{f(x)}\mathcal{O}f(x)\,,
\end{equation}
which is an elliptic function with periods \(2\mathbb{K}\) and \(2i\mathbb{K}'\) and a certain number of poles \(x_i\) of order \(p_i\) in the 
period-parallelogram. In terms of $F$, the eigenvalue equation becomes
\begin{equation}
 F(x)=\Lambda\,,
\end{equation}
and if $f(x)$ in (\ref{eq:ansatz}) is a solution of the differential equation, then the elliptic function \(F(x)\) should merely be a constant.
Therefore, we have to impose that in the Laurent expansion of \(F(x)\) 
\begin{equation}
\label{Fexpanded}
 F(\varepsilon+x_i)=\frac{A_{i,p_i}}{\varepsilon^{p_i}}+\frac{A_{i,p_i-1}}{\varepsilon^{p_i-1}}+...+\frac{A_{i,1}}{\varepsilon}+a_{i,0}+a_{i,1}
 \varepsilon+...
\end{equation}
all coefficients \(A_{i,j}\) of the principal part vanish.
This will constrain the free parameters in \eqref{potentials} and deliver the corresponding Bethe-ansatz equations for the spectral parameters \(\bar\alpha_i\).
%
\subsection{Pole structure}
\label{polestructure}
In order to proceed we need  to collect information about the pole structure of the functions appearing in 
\eqref{eq:ansatz}-\eqref{F}. In the study of their analytic properties, it is useful to introduce yet another function
\begin{equation} \label{Phi}
 \Phi(x)\equiv\frac{1}{f}\frac{\mathrm{d}f}{\mathrm{d}x}=\sum_{r=1}^n\left[Z(x+\bar\alpha_r+i\mathbb{K}')-Z(x)\right]+\rho+\lambda+
 \frac{n\pi i}{2\mathbb{K}}\,,
\end{equation}
which has $n+1$ poles at $x=i\KK'$ and $x=-\bar\alpha_1,-\bar\alpha_2,\dots\,, 
-\bar\alpha_n$, up to translations by the periods $2\KK$ and $2i\KK'$. We separately examine these two cases.\\
%

\subsubsection*{Expansion around the pole \(x=i\mathbb{K}'\)}

The expansion of the auxiliary function $\Phi$ (\ref{Phi}) around this singular point provides
 \begin{eqnarray}\label{eq:phiexpansionpole1}
&& \Phi(\varepsilon+i\mathbb{K}')=\frac{A_1}{\varepsilon}+a_0+a_1\varepsilon+a_2\varepsilon^2+a_3\varepsilon^3+\dots,  
\quad
 \Phi'(\varepsilon+i\mathbb{K}')=-\frac{A_1}{\varepsilon^2}+a_1+2a_2\varepsilon+3a_3\varepsilon^2+\dots\,, 
 \nonumber\\
&& \Phi''(\varepsilon+i\mathbb{K}')=\frac{2A_1}{\varepsilon^3}+2a_2+6a_3 \varepsilon+\dots\,, 
\qquad\Phi'''(\varepsilon+i\mathbb{K}')=-\frac{6A_1}{\varepsilon^4}+6a_3+\dots\,, 
\end{eqnarray}
where we denoted
\begin{eqnarray}
 A_1&=&-n,\qquad
 a_0=\sum_{r=1}^nZ(\bar\alpha_r)+\rho+\lambda, \qquad
 a_1=\frac{n}{3}(1+k^2)-k^2\sum_{r=1}^n\mathrm{sn}^2(\bar\alpha_r),\nonumber\\
 a_2&=&-k^2\sum_{r=1}^n\mathrm{sn}(\bar\alpha_r)\mathrm{cn}(\bar\alpha_r)\mathrm{dn}(\bar\alpha_r),\nonumber\\
 a_3 &=&\frac{n}{45}(1-16k^2+k^4)+\frac{2}{3}(1+k^2)k^2\sum_{r=1}^n\mathrm{sn}^2(\bar\alpha_r)-k^4\sum_{r=1}^n\mathrm{sn}^4(\bar\alpha_r).
\end{eqnarray}
The same procedure applied on the potentials (\ref{potentials}) leads to the series
\begin{eqnarray}\label{eq:pexpansionpole1}
 v_1(\varepsilon+i\mathbb{K}')&=&\frac{\alpha_1}{\varepsilon^2}+\alpha_0+\frac{\alpha_1}{3}(1+k^2)+\frac{\alpha_1}{15}(1-k^2+k^4)\varepsilon^2+0
 \cdot\varepsilon^3+...\\\nn
 v_2(\varepsilon+i\mathbb{K}')&=&-\frac{2\beta_2}{\varepsilon^3}+\frac{\beta_1}{\varepsilon^2}+\beta_0+\frac{\beta_1}{3}(1+k^2)+
 \frac{2\beta_2}{15}(1-k^2+k^4)\varepsilon+\frac{\beta_1}{15}(1-k^2+k^4)\varepsilon^2+...\nonumber\\\nn
 v_3(\varepsilon+i\mathbb{K}')&=&\frac{6\gamma_3}{\varepsilon^4}-\frac{2\gamma_2}{\varepsilon^3}+\frac{\gamma_1}{\varepsilon^2}+\gamma_0+
 \frac{\gamma_1}{3}(1+k^2)+\frac{2\gamma_3}{15}(1-k^2+k^4)+\frac{2\gamma_2}{15}(1-k^2+k^4)\varepsilon+\dots\,.
\end{eqnarray}

\subsubsection*{Expansion around the poles \(x=-\bar\alpha_i,\,i=1,...,n\)}

The analysis carried out for the family of poles $-\bar\alpha_i$ yields
\begin{equation} \label{eq:phiexpansionpole2}
 \Phi(\varepsilon-\bar\alpha_i)=\frac{1}{\varepsilon}+b_{0,i}+b_{1,i}\varepsilon+b_{2,i}\varepsilon^2
\end{equation}
where we identify the $\varepsilon$-coefficients with
\begin{eqnarray}
 b_{0,i}&=&\sum_{r\neq i=1}^nZ(\bar\alpha_r-\bar\alpha_i+i\mathbb{K}')+nZ(\bar\alpha_i)+\frac{i\pi(n-1)}{2\mathbb{K}}+\rho+\lambda,\nn\\
 b_{1,i}&=&-\sum_{r\neq i=1}^n\mathrm{cs}^2(\bar\alpha_i-\bar\alpha_r)+\frac{1}{3}(2-k^2)-n\mathrm{dn}^2(\bar\alpha_i),\nn\\
 b_{2,i}&=&-\sum_{r\neq i=1}^n\frac{\mathrm{cn}(\bar\alpha_i-\bar\alpha_r)\mathrm{dn}(\bar\alpha_i-\bar\alpha_r)}{\mathrm{sn}^3
 (\bar\alpha_i-\bar\alpha_r)}-nk^2\mathrm{sn}(\bar\alpha_i)\mathrm{cn}(\bar\alpha_i)\mathrm{dn}(\bar\alpha_i).
\end{eqnarray}
The potentials are regular functions.

\subsection{Consistency equations}

From the behaviour of $\Phi$ (\ref{Phi}) and the potentials (\ref{potentials}) around the singularities, it is now possible to reconstruct the pole 
structure of $F$, since the differential operators in (\ref{F}) translate into combinations of the auxiliary function and its 
derivatives:
\begin{eqnarray}\label{eq:derivatives}
 \frac{1}{f}\frac{\mathrm{d}^2f}{\mathrm{d}x^2}&=&\Phi(x)^2+\Phi'(x),\nonumber\\
 \frac{1}{f}\frac{\mathrm{d}^3f}{\mathrm{d}x^3}&=&\Phi(x)^3+3\Phi(x)\Phi'(x)+\Phi''(x),\nonumber\\
 \frac{1}{f}\frac{\mathrm{d}^4f}{\mathrm{d}x^4}&=&\Phi(x)^4+6\Phi(x)^2\Phi'(x)+4\Phi(x)\Phi''(x)+3\Phi'(x)^2+\Phi'''(x).
\end{eqnarray}
The condition of vanishing Laurent coefficients of $F$ at \(x=i\mathbb{K}'\) gives constraining equations on the numerical parameters \(\alpha_i,\beta_i\)
 and \(\gamma_i\), provided we take into account (\ref{eq:phiexpansionpole1})-(\ref{eq:pexpansionpole1}):
 {\small\begin{eqnarray}\nonumber
 \rho&=&-\sum_{r=1}^nZ(\bar\alpha_r)  ,\\\nonumber
0&=& n(n+1)(n+2)(n+3)+n(n+1)\alpha_1+2n\beta_2+6\gamma_3 ,\nonumber\\\label{eq:conditions}
0&=& \lambda[4(n+2)(n+1)n+2(n\alpha_1+\beta_2)]+(n\beta_1+2\gamma_2)\,, \\\nonumber
0&=& \lambda^2[6n(n+1)+\alpha_1]+\lambda\beta_1+a_1[-2n(n+1)(2n+1)+\alpha_1(1-2n)-2\beta_2]
\\ \nn &&+n(n+1)(\alpha_0+\frac{1}{3}\alpha_1(1+k^2))+\gamma_1\,,
 \\ \nonumber
0&=& 4\lambda^3n-2\lambda[a_1(6n^2+\alpha_1)-n(\alpha_0+\frac{\alpha_1}{3}(1+k^2))]-a_1\beta_1+2a_2[2n(1+n^2)+\alpha_1(n-1)+\beta_2]+\\\nonumber
&&+n(\beta_0+\frac{\beta_1}{3}(1+k^2))\,.\nonumber
\end{eqnarray}}
In particular, the term of~\(\mathcal{O}(\varepsilon^0)\) in the Laurent expansion gives the relation between the eigenvalue parameter $\Lambda$ and
the spectral parameters \(\bar\alpha_i\):
{\small\begin{eqnarray}\label{eq:spectralrelation}
\Lambda&=&\lambda^4+\lambda^2[6a_1(1-2n)+(\alpha_0+\frac{\alpha_1}{3}(1+k^2))]+\lambda\,[\,a_2(4(2+3n(n-1))+2\alpha_1)+(\beta_0+\frac{\beta_1}{3}(1+k^2))]+ \nonumber\\\nonumber
 &&+a_1^2[3(1-2n(1-n))+\alpha_1]+a_1(1-2n)(\alpha_0+\frac{\alpha_1}{3}(1+k^2))+a_2\beta_1+ \\\nonumber
&& +a_3[2(2n-1)(n(1-n)-3)+\alpha_1(3-2n)-2\beta_2]+\\
&&+\frac{1}{15}(1-k^2+k^4)[n(n+1)\alpha_1-2n\beta_2+2\gamma_3]+\gamma_0+\frac{\gamma_1}{3}(1+k^2) ~.
 \end{eqnarray}}
Finally, imposing that the $1/\varepsilon$-coefficient around the poles \(x=-\bar\alpha_i\) should vanish gives the
Bethe-ansatz equations for the spectral parameters ($i=1,...,n$)
\begin{equation}\label{eq:Betherel}
 4b_{0,i}^3+2b_{0,i}(6b_{1,i}+\alpha_0+\alpha_1k^2\mathrm{sn}^2(\bar\alpha_i))+8b_{2,i}+\beta_0+\beta_1k^2\mathrm{sn}^2(\alpha_i)-
 2\beta_2k^2\mathrm{sn}(\bar\alpha_i)\mathrm{cn}(\bar\alpha_i)\mathrm{dn}(\bar\alpha_i)=0~,
\end{equation}
where we used (\ref{eq:phiexpansionpole2}). In deriving these conditions, we have assumed that \(\alpha_i\neq\alpha_j\) for any \(i,j=1,...,n\).

It is important to mention that in all the examples successfully analyzed in this paper  we have made use 
of the $n=1$ consistency equations alone - therefore using a single factor in the product 
defining \eqref{eq:ansatz} - which we report here separately for reader's convenience.

\subsubsection*{$n=1$ consistency equations}

\begin{eqnarray}\label{eq:n1conditions}
0&=&12+\alpha_1+\beta_2+3\gamma_3, \\\nonumber
0&=& \lambda(24+2(\alpha_1+\beta_2))+\beta_1+2\gamma_2\,, \\\nonumber
0&=& \lambda^2(12+\alpha_1)-a_1(\alpha_1+12+2\beta_2)+\lambda\beta_1+2\left(\alpha_0
+\frac{1}{3}\alpha_1(1+k^2)\right)+\gamma_1\,, \\\nonumber
 0&=&-4\lambda^3-2\lambda\left(\alpha_0-a_1(6+\alpha_1)+\frac{1}{3}\alpha_1(1+k^2)\right)+a_1\beta_1
 -2a_2(4+\beta_2)-\beta_0-\frac{1}{3}\beta_1(1+k^2)\,,\\ \nn
\Lambda&=& \lambda^4+\lambda^2\,[\,-6a_1+\alpha_0+\frac{1}{3}\alpha_1(1+k^2)\,]
+\lambda\,[\,2a_2(\alpha_1+4)+\beta_0+\frac{1}{3}\beta_1(1+k^2)\,]+\\\nonumber
&+& a_1^2(\alpha_1+3)-a_1(\alpha_0+\frac{1}{3}\alpha_1(1+k^2))+a_2\beta_1+a_3(\alpha_1-6-2\beta_2)+\\\nonumber
&+&\frac{2}{15}(1-k^2+k^4)(\alpha_1-\beta_2+\gamma_3) 
 +\gamma_0+\frac{1}{3}\gamma_1(1+k^2)~.
\end{eqnarray}
For $n=1$, the condition for the pole at \(x=-\bar\alpha \) to vanish 
turns out to be equivalent to the fourth equation in (\ref{eq:n1conditions}) and therefore it does not give any further constraint.

Our result for the consistency equations is in partial disagreement with the study in~\cite{Mitt}. 
However, the examples discussed in the next section and in Appendix \ref{app:squaredlame}, 
as well as numerical cross checks of the provided solutions,  give strong evidence for the correctness of our procedure.

\section{Exact bosonic one-loop partition functions for folded string}
\label{sec:exactpartition}

We are now ready to use the analysis performed in Section~\ref{sec:4thorder} for 
the  computation of determinants of the fluctuation operators
discussed in Section \ref{sec:generalfolded}.

\subsection{Exact partition function and one-loop energy for the LL folded string}
\label{sec:exactpartLL}
  
\begin{figure}
   \centering
 \includegraphics[scale=0.4]{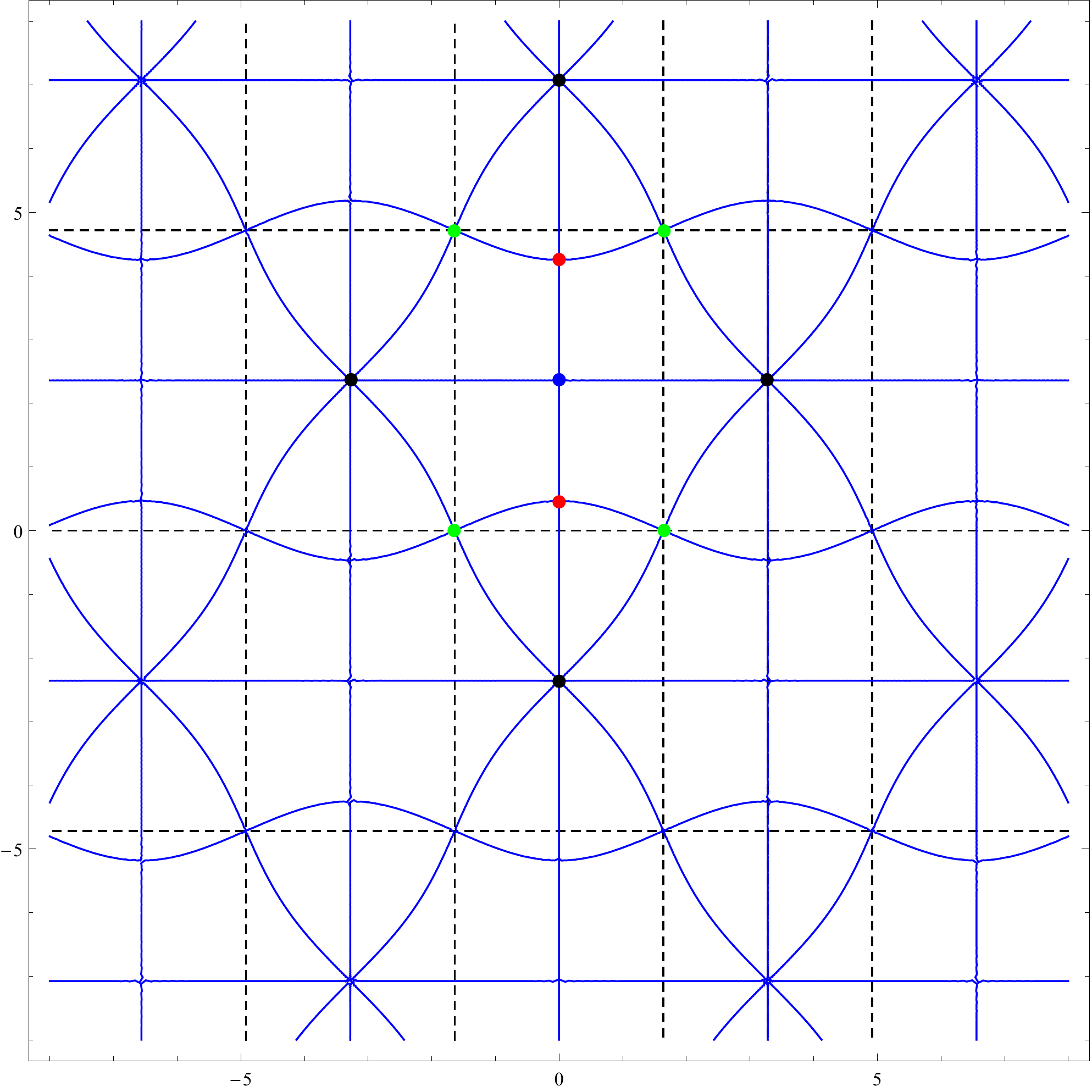}
 \caption{In the complex \(\alpha\) plane one can plot the lines where \(\Omega^2(\alpha)\) is real. 
 The ``physical'' four linear independent solution live on these lines. 
 The green dots represent places where \(\Omega^2=0\), 
 red for \(\Omega^2=4k^2k'^2\), blue for \(\Omega^2=1\) and black for poles. We have chosen $k=0.4$.}
 \label{fig:spectralplane}
\end{figure}

\begin{figure}
   \centering
 \includegraphics[scale=0.6]{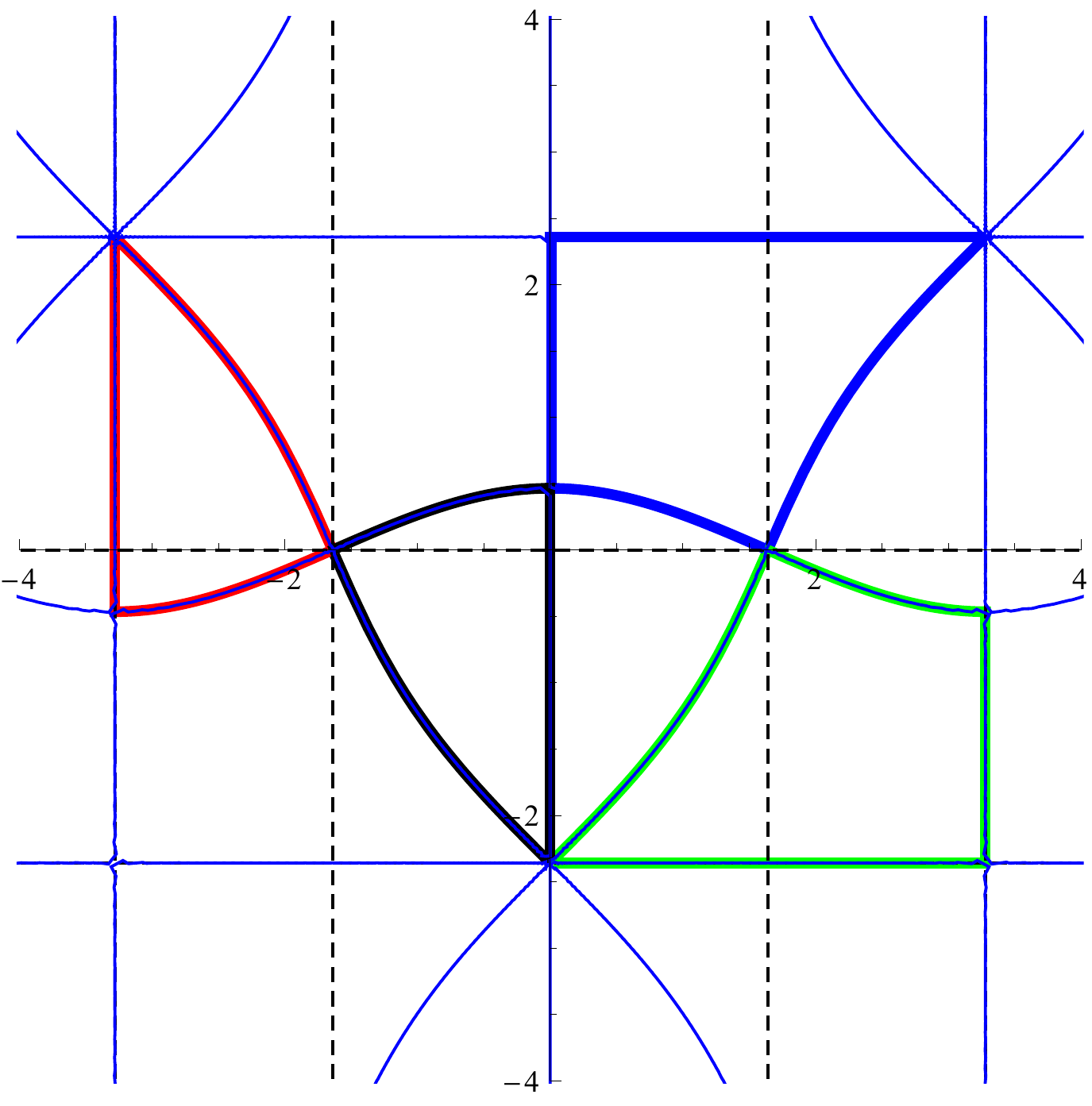}
 \caption{In the fundamental domain the four independent solutions of the fourth order differential 
 operator \eqref{LL_first}  are marked with colors. 
 Walking on such a closed path, \(\Omega^2\) runs from \(-\infty\) to
 \(+\infty\). We have chosen $k=0.4$.}
 \label{fig:fundomain}
\end{figure}

The fourth order differential operator in \eqref{LL_first}, governing the fluctuations of the LL quantum 
model defined by \eqref{L_LL}-\eqref{OLL}, is easily seen  to be of the type (\ref{Oholy}) once the 
following identification is performed 
\ba\nonumber
 \alpha_0&=&2(1+2k^2),\qquad \alpha_1=-8,\qquad\qquad \beta_0=\beta_1=0,\qquad \beta_2=-4\,,\\
 \gamma_0&=&1-\frac{\omega^2}{4\mathrm{w}^2},
 \qquad ~\gamma_1=\gamma_2=\gamma_3=0\,.
\ea
Using the \(n=1\)  consistency equations \eqref{eq:n1conditions}, one finds  
(here $\bar\alpha=\alpha$)
\begin{equation}
 \lambda=\pm k\sqrt{\mathrm{sn}^2(\alpha)-1}~,
\end{equation}
where the relation between \(\Omega\) and \(\alpha\) is
\begin{eqnarray}\label{eq:spectralLL}
 \Omega^2_{\mp}(\alpha)&=&4k^2-4k^2(1+2k^2)\mathrm{sn}^2(\alpha)+8k^4\mathrm{sn}^4(\alpha)\mp
 8k^3\mathrm{sn}(\alpha)\mathrm{cn}(\alpha)\mathrm{dn}(\alpha)\sqrt{\mathrm{sn}^2(\alpha)-1}\nonumber\\
 &=& 
 4k^2\mathrm{cn}^2(\alpha)\left[ik\,\mathrm{sn}(\alpha)\mp\mathrm{dn}(\alpha)\right]^2~.
\end{eqnarray}
It seems advantageous to consider \(\alpha\in\mathbb{C}\) as the
independent parameter and therefore \(\Omega\) as a doubly periodic function of \(\alpha\) as in (\ref{eq:spectralLL}).
 There should exist four values of \(\alpha\), which correspond to one value of \(\Omega\). To be more precise, for
the \emph{physical spectrum} we are looking for all values of \(\alpha\), which correspond to a \emph{real \(\Omega^2\)}.
The analysis of Appendix \ref{sec:spectraldomain} is devoted to this study, and is nicely summarized in Fig. 
\ref{fig:spectralplane} where in the complex \(\alpha\) plane the lines where \(\Omega^2(\alpha)\) is real 
are plotted.  
The ``physical'' four linear independent solutions of the  fourth order differential operator
\eqref{eq:4thorder} live on these lines, and in the fundamental domain 
represented in Fig. \ref{fig:fundomain} they correspond to the different 
colours. 
In the following let \(0<k<1/\sqrt{2}\). The case \(1/\sqrt{2}<k<1\) can be obtained by applying the 
duality transformation of  Appendix \ref{app:duality} on the following
results.

For a given real value of \(\Omega^2\) the four independent solutions read
\begin{equation}
 f_i(x,\Omega,k)=\frac{H(x+\alpha_i)}{\Theta(x)}e^{-x\left[Z(\alpha_i)-ik\mathrm{cn}(\alpha_i)\right]},\qquad i=1,\dots\,,4\,,
\end{equation}
where the \(\alpha_i\) as function of \(\Omega\) have to be chosen according to the range of 
$\Omega$. For example,  for \(-\infty<\Omega^2<0\), one has
\begin{eqnarray}\label{alphabranchneg}
 \alpha_1(\Omega,k)&=&u(\Omega,k)-iv(\Omega,k),\\ \nn
 \alpha_2(\Omega,k) &=& 2\mathbb{K}-u(\Omega,k)+iv(\Omega,k),\\ \nn
\alpha_3(\Omega,k)&=&2\mathbb{K}+u(\Omega,k)+iv(\Omega,k),\\ \nn
 \alpha_4(\Omega,k) &=& 2\mathbb{K}-u(\Omega,k)+2i\mathbb{K}'-iv(\Omega,k),
\end{eqnarray}
where
\begin{equation}
 u(\Omega,k)=\mathrm{sn}^{-1}\Big[\textstyle{\sqrt{\frac{2}{\Omega^2}(1-\sqrt{1-\Omega^2})}},\,k\Big],\qquad 
 v(\Omega,k)=\mathrm{sn}^{-1}\Big[\sqrt{\frac{\Omega^2-2k^2+2k^2\sqrt{1-\Omega^2}}{\Omega^2-4k^2k'^2}},\,k'\Big]~.
\end{equation}
The expressions for the $\alpha_i$'s in the other ranges of $\Omega$ are 
collected in \eqref{alphabranchpos1}-\eqref{alphabranchpos3}.
The quasi-momenta \(p_i\) are obtained  by \(f_i(x+2\mathbb{K})=e^{2\mathbb{K}ip_i}f_i(x)\) as
\begin{equation}\label{quasimomentaLL}
 p_i(\Omega,k)=iZ(\alpha_i,k)+k\mathrm{cn}(\alpha_i,k)-\frac{\pi}{2\mathbb{K}},
\end{equation}
where the corresponding \(\alpha_i(\Omega,k)\) have to be chosen according to the previous list.
By construction~\footnote{See discussion around \eqref{eq:conditionFloquet}-\eqref{eq:Floquetfactors}.},  
 only two of the quasi-momenta are independent, which we will call $p_1$ and $p_2$.

The \emph{exact} determinant of the Landau-Lifshitz model defined by \eqref{L_LL}-\eqref{OLL}, using \eqref{detop} with $2L=4\KK$,  reads then 
\begin{eqnarray}\label{detLL}
\det{\mathcal O_{LL}}&=& 16 \sin^2\big(2\KK\,p_1(\Om,k 
)\big)\,\sin^2\big(2\KK\,p_2(\Om,k)\big)~.
\end{eqnarray}


We can  immediately recover the characteristic frequencies of the problem, found in~\cite{Minahan:2005mx} using operator methods up to second-order perturbation theory, 
by simply looking at the   zeroes of the determinant \eqref{detLL}, where 
the quasi momenta are built with the $\alpha$'s in the  branch $\Omega^2>0$ and 
are Taylor-expanded around $k=0$. It is enough to look to the factor in \eqref{detLL} involving $p_1$, whose expansion 
re-expressed in terms of the $\omega$ is 
\begin{eqnarray}\nonumber
p_{1}&=&\sqrt{2 \omega +1}  -\frac{ \omega }{2 \sqrt{2 \omega +1}}k^2 +\frac{ \left(-10 \omega ^4-11 \omega ^3+6 \omega +2\right)}
  {32 \omega ^2 (2 \omega +1)^{3/2}}k^4
\\\label{pexpfrequencies}
&&+
\frac{\left(-22 \omega ^7-39 \omega ^6-19 \omega ^5+2 \omega ^4+10 \omega ^3
+16 \omega ^2+10 \omega +2\right)}{64 \omega ^4 (2 \omega +1)^{5/2}}k^6  +O(k^8)~.
\end{eqnarray}
Inserting it into \eqref{detLL} and requiring the vanishing of the expression order by order in small $k^2$, one finds the (squared) frequencies to 
be
\begin{equation}
\omega^2=\frac{1}{4}
   (n^2-1)^2+
  \frac{1}{4} (1-n^2)k^2+\frac{
   \left(3 n^4-2
   n^2+15\right)}{64(1-n^2)}k^4-\frac{
   \left(n^8+n^6+7 n^4+27
   n^2+28\right)}{128
   \left(n^2-1\right)^3}k^6+O\left(k^8\right)~,
\end{equation}
where we do not report higher orders, but notice that it is straightforward to calculate them. 
The first three orders of the expansion above coincide with the ones of~\cite{Minahan:2005mx}.

\bigskip

The one-loop correction to the $SU(2)$ LL string energy can be of course obtained perturbatively via a 
regularized sum over the frequencies given above~\cite{Minahan:2005mx}, or exactly  
 in terms of the one-loop world-sheet effective action $\Gamma^{(1)}$, 
and thus in terms of the corresponding partition function $Z_{LL}$,  as follows
\be
E_1= {\Gamma^{(1)}\over \mathcal T}= -{\log Z_{LL} \over \mathcal T}\,, \qquad\quad  \mathcal T=\int_{-\infty}^\infty d\tau \,.
\ee
The Euclidean LL partition function is obtained from the functional determinant as
\be
\label{part_function_LL}
Z_{LL}= {\det}^{-1/2} \mathcal O_{LL} \,,
\ee
which using \eqref{detLL} can be explicitly written as~\footnote{It is convenient to use as integration variable the rescaled frequency
 $\Omega$, see \eqref{Omega}. }  
\be\label{Gamma}
\Gamma^{(1)}= -\log Z_{LL}
={\mathcal T\over 2} \int^{\infty}_{-\infty}{d\Omega\over 2\pi} \log  \left[
16 \sin^2\big(2\KK\,p_1(\Omega, k 
)\big)\,\sin^2\big(2\KK\,p_2(\Omega,k)\big)\right]\,.
\ee
Above, the Euclidean setting requires the quasi-momenta $p_i$ to be built out of the $\alpha$'s 
in the branch $\Omega^2<0$, which are given in \eqref{alphabranchneg}. 
The  integral in \eqref{Gamma} is divergent. A first meaningful choice of regularization of the functional determinant 
is to refer  it to the $k=0$ case. Indeed this limit, as discussed in~\cite{Minahan:2005mx}, 
represent a nearly point-like string and the correction to the ground-state energy should vanish. 
Hence, we obtain 
\be\label{Gamma_reg}
\Gamma^{(1)}_{reg}  ={\mathcal T\over 2}  \int^{\infty}_{-\infty}{d\Omega\over 2\pi} \log\left[ 
 \sin^2\big(2\KK\,p_1(\Omega,k)\big)\,\sin^2\big(2\KK\,p_2(\Omega,k)\big)
\over  \sin^2\big(\pi\,p_1(\Omega,0)\big)\,\sin^2\big(\pi\,p_2(\Omega,0)\big) \right]\,
\ee
where, at the denominator, the quasi-momenta $p_i(\Omega,0)$ are computed at $k=0$.
 In order to analytically perform the above integral over $\Omega$, 
we can resort to the short string expansion $k^2\simeq 0$. 
This again means to consider the small $k$ expansion of quasi-momenta $p_i$ 
 (which differ from the ones considered above, as we are in a different branch for the $\alpha$'s),  
 as reported in   Appendix \ref{app:shortstring}, and then after, to integrate over $\Omega$ the 
 corresponding expressions computed order by order in $k$. 
Each term in the $k^2$-series for $\Gamma^{(1)}_{reg}$ must be further regularized, which is of course 
expected as only certain bosonic degrees of freedom and no fermionic ones (crucial for UV finiteness) 
participate to the effective LL action. In  Appendix \ref{app:shortstring} we report two different ways of 
regularizing (one inspired by $\zeta$-function regularization and one with standard  cutoff) which 
lead to the same result.
%
%
The resulting expression for the $k^2$-expansion of the one-loop energy is the 
same as in~\cite{Minahan:2005mx} 
\be
\label{one_loop_energy}
E_1={\Gamma^{(1)}_{reg}\over \mathcal T}= {1\over 4} k^2+ {1\over 16}\left(1-{\pi^2\over 3}\right) k^4+O(k^5)\,.
\ee
It is interesting to notice that this result follows smoothly by our standard
regularization of the 2d LL string effective action, while in~\cite{Minahan:2005mx} 
it is implied by a $\zeta$-function regularization supplemented by a general prescription 
for the vacuum energy in terms of characteristic frequencies of a mixed system of oscillators~\cite{Blau:2003rt}.

From equation \eqref{useful_LL}, in terms of the physical parameter $J_2/J$, the short string limit $k^2\to 0$ reads
\be
{J_2\over J}= {k^2\over 2} + {k^4\over 16}+ \O(k^5) \,, \qquad k^2 =  {2\, J_2\over J}- \half \left({J_2\over J}\right)^2\,,
\ee
and the expression for the  energy becomes
\be
E_1 = {\tilde \lambda\over 2}\left({J_2\over J}+ \left({1\over 4}-{\pi^2\over 6}\right)\left({J_2\over J}\right)^2\right) +O\left(\left({J_2\over J}\right)^2\right)\,,
\ee
where we restored the $\tilde\lambda$ dependence.  The first three terms in the formula above are in agreement with \cite{Minahan:2005mx}.

\bigskip

For completeness, we mention that the analysis for the LL folded string in the $\grSL(2)$ sector 
(where strings rotate in $\ads_3 \subset\ads_5$ with center of mass moving along a big circle of 
$S^5$) is totally analogous. Using 
the following analytical continuation\cite{Minahan:2005mx,Beisert:2003ea}
\[
\psi\rightarrow - i \r\,, \qquad \vf\rightarrow \eta\,, \qquad w_1\rightarrow \kappa \,,\qquad w_2 \rightarrow w_1\,,
\]
one can easily see that the system of coupled fluctuations is effectively described 
by the fourth order differential operator in \eqref{LL_first} where now $k^2$  is negative.
Its solutions are then trivially generalised to the case $k^2<0$ as basically in each formula 
one should substitute $k\to\sqrt{-k^2}$ and omit all imaginary constants $i$  in the 
exponentials.  

 \subsection{Folded string in full bosonic sigma-model}

\begin{figure}
 \centering
 \includegraphics[scale=0.5]{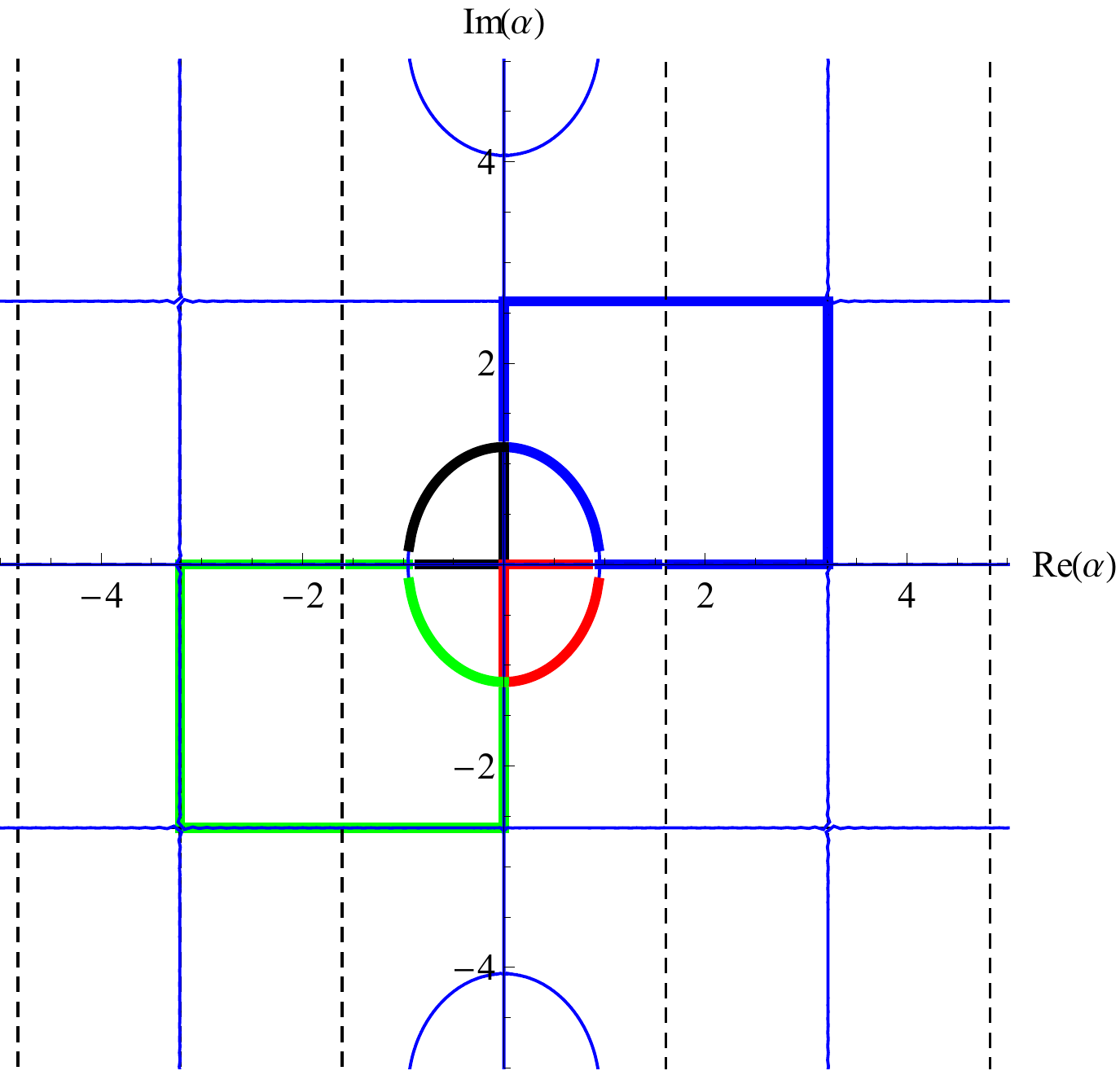}
 \caption{The places, in the fundamental domain of the complex \(\alpha\) plane, where 
 the four linear independent solutions \eqref{eq:foursolutionsfolded2}-\eqref{eq:foursolutionsfolded2bis} (marked with different colours) for the 
 fourth order differential operator \eqref{eq:opnunonzero2} live. Here we have chosen $k=3,w=6,\nu=2.5$. See also Fig. \ref{fig:FSPlane1}. }
\label{fig:fonddomain2}
\end{figure}

The fourth order differential operator in \eqref{eq:opnunonzero2}~\footnote{In this section we are working in Minkowski signature, so that 
\eqref{foldedcoeff} are obtained from \eqref{Oholy} analytically continuing the frequencies.} is again of the type (\ref{Oholy}) with the identification 
\ba\label{foldedcoeff}\nonumber
 \alpha_0 &=& 2(\bar\Omega^2+k^2+1),\qquad \alpha_1=-8,\qquad\qquad \beta_0=\beta_1=0,
 \qquad \beta_2=-4\,\\
 \gamma_0 &=& [(-\bar\Omega^2+1+k^2)^2-4k^2]-\frac{4\nu^2\bar\Omega^2}{\bar w^2-\nu^2},
 \qquad ~\gamma_1=\gamma_2=\gamma_3=0\,,
\ea
where $\bar \Omega$ is defined in \eqref{baromega}. 
Using the consistency equations \eqref{eq:n1conditions} one finds 
\begin{equation}
 \lambda=\pm\sqrt{k^2\mathrm{sn}^2(\alpha)-\bar\Omega^2}~,
\end{equation}
where the relation between \(\bar\Omega\) and \(\alpha\) is
\begin{equation}
 8k^4\mathrm{sn}^4(\alpha)-4(1+k^2+\bar\Omega^2)k^2\mathrm{sn}^2(\alpha)\pm
8k^2\mathrm{sn}(\alpha)\mathrm{cn}(\alpha) \mathrm{dn}(\alpha)\textstyle\sqrt{k^2\mathrm{sn}^2(\alpha)-\bar\Omega^2}
-\textstyle\frac{4\nu^2\bar\Omega^2}{\bar w^2-\nu^2}=0\,. 
\end{equation}
The study reported in Appendix \ref{sec:spectraldomain_folded} shows that the ``physical''  four linear 
independent solutions of \eqref{eq:opnunonzero2} live on the straight and ellipse-like lines 
in Fig.~\ref{fig:fonddomain2}.

Using in  \(f_i(x+2\mathbb{K})=e^{2\mathbb{K}ip_i}f_i(x)\) their explicit expressions 
- cf.  \eqref{eq:foursolutionsfolded2} - the quasi-momenta are then obtained as
\begin{equation}\label{quasimomnu}
 p_n(\bar\Omega)=\pm i\left[Z(\alpha_n)+\frac{k\,\mathrm{sn}(\alpha_n)}{(\kappa^2-
 \nu^2)\mathrm{sn}^2(\alpha_n)+\nu^2}\left(\kappa\bar w\pm\sqrt{\kappa^2-\nu^2}\sqrt{\bar w^2-\nu^2}\,
 \mathrm{cn}(\alpha_n)\mathrm{dn}(\alpha_n)\right)\right]+\frac{\pi}{2\mathbb{K}},
\end{equation}
where \(\alpha_n\) as function of \(\bar\Omega\) has to be chosen from the list 
in \eqref{alphafolded1}-\eqref{alphafoldedfin}.
The functional determinant is again given by
\begin{equation}\label{detbosfolded}
 \mathrm{det}\,\mathcal{O}_{\nu}=16\sin^2({L}\,p_1)\sin^2(L\,p_2)~,
\end{equation}
with $2L=4\KK$. 

As a first check of the correctness of the procedure, one can take the long string limit 
\(k\to 1, ~\bar w^2\to\kappa^2\) (see Section \ref{app:foldedfullstatic}) and look  at the zeroes of the expression 
above choosing the positive-frequency range \eqref{C32} and \eqref{alphafoldedfin} for the 
$\alpha$'s in \eqref{quasimomnu}. We obtain the characteristic frequencies~\footnote{In this limit, 
we obtain, from \eqref{C32} and \eqref{alphafoldedfin},
\begin{equation}
 \mathrm{sn}(\alpha_{1,2})\to -\nu^2\omega/\Big[\sqrt{\kappa^2-\nu^2}\sqrt{\left(\kappa^2\pm
 \sqrt{(\kappa^2-\nu^2)^2+\nu^2\omega^2}\right)^2-\nu^4}\Big]~,
\end{equation}
where we have used \eqref{baromega}.}
 \begin{equation}\label{frequencylong}
 \omega_n=\sqrt{n^2+2\kappa^2\pm2\sqrt{\kappa^4+n^2\nu^2}},
\end{equation}
which is the same result as found in \cite{Frolov:2006qe}~\footnote{ 
One can obtain this result also from (\ref{eq:opnunonzero2}), which in this limit becomes 
\begin{equation} 
\lim_{k\to 1}\mathcal O^{(4)} = \partial_x^4+2[\omega^2-2(\kappa^2-\nu^2)]\partial_x^2+(\omega^4-4\omega^2\kappa^2)~.
\end{equation}}, see also \eqref{goodfreq}.

\bigskip

Since we are missing (see Introduction and Section \ref{sec:foldedfermions}) the fermionic 
counterpart of \eqref{detbosfolded}, we cannot proceed with the exact evaluation 
of the full (superstring) one-loop partition function on the folded two-spin solution. 
However, we observe that a nice consequence of our  procedure is the 
possibility of making a non-trivial, analytical statement on the equivalence of 
partition functions
in conformal and static gauge in the single-spin ($\nu=0$) case. 
While here the  fermionic determinant can be  given exactly for all values of the 
spin~\cite{Beccaria:2010ry}, it is only the bosonic partition function in static gauge 
- where fluctuations are naturally decoupled - which 
has been written down in an analytically exact closed form, and 
reads~\cite{Beccaria:2010ry} 
\begin{eqnarray}\label{Zbosstatuc}
\log Z_{\rm static~gauge}^{\rm bos}&=-&\frac{{\cal T}}{2}\int\frac{d\omega}{2\pi}\log\Big(\det\mathcal{O}_{\phi}\,{\det}^2\mathcal{O}_{\beta} 
{\det}^5 \mathcal{O}_{0} \Big)
\end{eqnarray}
where
\begin{eqnarray}\label{detstatic}
\!\!\!\!\!\!\!\!\!\!\!  \!\!\!
\det\mathcal{O}_{\phi}=4\sinh^2[2\tilde{\mathbb{K}}Z(\alpha_{\phi}|\tilde 
  k^2)]\,,~~
  \det\mathcal{O}_{\beta}=4\sinh^2\left[2\mathbb{K}\,Z(\alpha_{\beta}|k^2)\right]\,,~~
  {\det} \mathcal{O}_{0}=4\sinh^2(\pi \omega)
\end{eqnarray}
and
\begin{equation}
 \!\!\!\!\!\! 
\mathrm{sn}(\alpha_{\phi}\,|\tilde k^2)
=\frac{1}{\tilde k}\sqrt{1+\left(\frac{\pi\omega}{2\tilde{\mathbb{K}}}\right)^2}\,
,~~
  \mathrm{sn}(\alpha_{\beta}|k^2)=\frac{1}{k}
 \sqrt{1+k^2+\left(\frac{\pi\omega}{2\mathbb{K}}\right)^2}~,
 \end{equation}     
with \(\tilde k^2=4k/(1+k)^2\) and  \(\tilde{\mathbb{K}}=\mathbb{K}(\tilde k^2)\).
The analysis  in Section \ref{sec:foldedfullbos} shows 
that, in conformal gauge, 
the spectral problem associated to the mixed-mode, $3\times 3$ matrix differential 
operator corresponding to \eqref{eom1}-\eqref{eom3}  can 
be evaluated, see \eqref{eq2}, via the 
product of a free determinant times the determinant of the fourth order 
differential operator \eqref{eq:opnunonzero2}, and thus~\footnote{While we worked 
at the operatorial level with the linearized (near folded string solution) form of the string equations of
motion, and did not prove the 
formal equivalence between the determinant of the $3\times 3$ matrix differential operator corresponding to 
\eqref{eom1}-\eqref{eom3}  and the product $\det\mathcal{O}_{\nu=0}\det\mathcal{O}_{0}$, 
  \eqref{Zbosconformal} should be formally correct. This is not different from the steps 
  \eqref{L_LL}-\eqref{OLL} followed in setting the LL spectral problem,  
  with a new ingredient here consisting in the implementation of Virasoro constraints. As thoroughly discussed in 
~\cite{Beccaria:2010ry},  at the level of path integral  the step analogous to 
\eqref{tfromVIR}-\eqref{phifromVIR} will produce an extra $\det\mathcal{O}_{0}$ factor as required for 
balance of degrees of freedom. }
\begin{eqnarray}\label{Zbosconformal}
\log Z_{\rm conformal~gauge}^{\rm bos}=-\frac{{\cal T}}{2}\int\frac{d\omega}{2\pi}\log\Big(
\det\mathcal{O}_{\nu=0}\,{\det}^2\mathcal{O}_{\beta}\,
{\det}^4 \mathcal{O}_{0} \Big)~,
\end{eqnarray}
where in the counting of massless operators we already have taken into account the two conformal 
gauge massless ghosts~\cite{Drukker:2000ep}, and (see Appendix \ref{app:nu0})
\begin{equation}\label{detonu0}
\det\mathcal{O}_{\nu=0}=16\sinh^2\left[2\mathbb{K}\left(Z(\alpha|k^2)+\frac{1+\mathrm{cn}(\alpha|k^2)
\mathrm{dn}(\alpha|k^2)}{\mathrm{sn}(\alpha|k^2)}\right)\right]\sinh^2(2\mathbb{K}\bar\Omega)~, 
\end{equation}
with (switching to Euclidean signature)
\begin{equation}
    \mathrm{sn}^2(\alpha|k^2)=\frac{-4\bar\Omega^2}{(1+k^2-\bar\Omega^2)^2-4k^2}~.
\end{equation}
One can see that the second factor in \eqref{detonu0} corresponds to the same massless boson mode of 
\eqref{detstatic} (recalling 
\eqref{baromega}
and that 
 for $\nu=0$ it is $\bar w=\frac{2\KK}{\pi}$), while for the first factor one should 
use for the Jacobi Zeta function the transformation \eqref{Landen}
which, writing
$\tilde\alpha= \alpha/(1+\tilde k')+ i\mathbb{K}'/(1+\tilde k')$, leads to the identity
\begin{equation}
 2\mathbb{K}\left[Z(\alpha|k^2)+\frac{1+\mathrm{cn}(\alpha|k^2)\mathrm{dn}(\alpha|k^2)}{\mathrm{sn}(\alpha|k^2)}\right]
 =2\mathbb{\tilde K}Z(\tilde\alpha|\tilde k^2)+i\pi~.
\end{equation}
This establishes  analytically the equivalence of static and conformal gauge 
bosonic determinants \eqref{Zbosstatuc}-\eqref{Zbosconformal}~\footnote{At the operator level, it was noticed already in~\cite{Beccaria:2010ry} that ${\mathcal O}_{\nu=0}$ 
manifestly factorizes as a product of two second-order ones
\begin{equation}
\label{fac}
 {\cal O}^{(4)}_{\nu=0}=  {\cal O}_1  \cdot  {\cal O}_2  \  ,  \quad
{\cal O}_1
 = (\rho')^{-1} \,\Big[ \partial^2_\sigma+\omega^2-2\,\rho'^2-2\frac{\kappa^2\,w^2 }{\rho'^2} 
 \Big]  \, \rho'   
 \ , \quad
  {\cal O}_2  =  {\rho'}\ \ [\partial^2_\sigma+\o^2] \  {(\rho')^{-1}} \ , 
\end{equation}
where the operators within brackets are those, decoupled, 
appearing in static gauge~\cite{Frolov:2002av}.}.

\section{Outlook}

In this paper we have made a first step into the analytic solution of the matrix 
fluctuations determinant for nontrivial string configurations relevant for the study of the 
AdS/CFT integrable systems, evaluating exactly the one-loop partition function 
for the quantum Landau-Lifshitz model on the $SU(2)$ folded string solution of~\cite{Minahan:2005mx}. 
The same procedure allows the diagonalization of the bosonic sector of fluctuations of 
the full $\ads_5\times \sphere^5$ excitations over the two-spin folded string solution 
of~\cite{Frolov:2002av}. 

This result calls for the complete (i.e. including fermions) 
solution of the fluctuation problem for non-homogeneous configurations of elliptic type, 
which might require a nontrivial field redefinition for the corresponding Lagrangian, or equivalently a modification 
of the ansatz for the solution of the related differential operator.
This class of solutions includes the  relevant case of  open string configurations corresponding to the 
space-like Wilson loops of~\cite{Drukker:2011za} 
(also in other backgrounds~\cite{Forini:2012bb}). 
Completing in this sense the analysis here performed should give an answer to the caveats of the 
semiclassical analysis mentioned in the Introduction,   enlarging the range of applicability 
of the procedure and opening the way to the detailed understanding of the relation 
between this quantum field-theoretical approach and the one based on the 
algebraic curve~\cite{Gromov:2013pga, Cavaglia:2014exa}.
 


\section*{Acknowledgments }

We are grateful to M.~Beccaria, G. Dunne and A. A. Tseytlin for earlier collaboration on the topic of Section 
\ref{sec:foldedfullbos} and \ref{app:foldedfullstatic}. 
It is a pleasure to thank M. ~Beccaria, G.~ Dunne, S.~ Frolov, L.~ Griguolo, 
D. ~Seminara, L.~Thorlacius and A.~A.~Tseytlin for discussions.  
 The work of V.F., M.P. and E.V.  is funded by the Emmy Noether Programme 
 ``Gauge Field from Strings'' funded by DFG. 
 M.P. also acknowledges support from SFB 647 "Space-Time-Matter.  
 Analytic and Geometric Structures".  
V.G.M.P. acknowledges partial support from Swedish Research Council 
for funding under the contract 
623-2011-1186.


\appendix
 
 \section{The squared Lam\'e operator}
 
\label{app:squaredlame}

As a check of the procedure described in Section \ref{sec:4thorder} and of the involved algebraic 
manipulations there performed,
we consider the fourth order differential operator  obtained by squaring the Lam\'e operator
\begin{equation}
 \mathcal{O}_L=-\partial_x^2+2k^2\mathrm{sn}^2(x|k^2)+\Omega^2,
\end{equation}
which gives
\begin{eqnarray}\label{eq:Lame4}
 \mathcal{O}_L^2&=&\partial_x^4-2(2k^2\mathrm{sn}^2(x)+\Omega^2)\partial_x^2
 -8k^2\mathrm{sn}(x)\mathrm{cn}(x)\mathrm{dn}(x)\partial_x\\ \nn
&& -8k^4\mathrm{sn}^4(x)+4(2(1+k^2)+\Omega^2)k^2\mathrm{sn}^2(x)-4k^2+\Omega^4~.
\end{eqnarray}
Since the solution of the Lam\'e equation is well known we can immediately write down the Floquet solutions for
\begin{equation}
 \mathcal{O}_L^2 f(x)=\Lambda f(x),
\end{equation}
given by
\begin{eqnarray}
 f_1(x)&=&\frac{H(x+\alpha_+)}{\Theta(x)}e^{-xZ(\alpha_+)},\qquad f_2(x)=\frac{H(x-\alpha_+)}{\Theta(x)}e^{xZ(\alpha_+)},\\\nn
 f_3(x)&=&\frac{H(x+\alpha_-)}{\Theta(x)}e^{-xZ(\alpha_-)},\qquad f_4(x)=\frac{H(x-\alpha_-)}{\Theta(x)}e^{-xZ(\alpha_-)}\,,
\end{eqnarray}
with
\begin{equation}\label{eq:solutionLame4}
 \mathrm{sn}(\alpha_{\pm}|k^2)=\sqrt{\frac{1+k^2\mp\sqrt{\Lambda}+\Omega^2}{k^2}}\,.
\end{equation}
For the squared Lam\'e operator (\ref{eq:Lame4}) we can read off the coefficients
\begin{eqnarray}
 \alpha_0&=&-2\Omega^2,\qquad\alpha_1=-4,\quad \beta_0=0,\qquad \beta_1=0,\qquad\beta_2=-4\\\nn
 \gamma_0&=&-\frac{4}{3}k^2+\Omega^4 ,\qquad\gamma_1=\frac{8}{3}(1+k^2)+4\Omega^2,
 \qquad \gamma_2=0,\qquad \gamma_3=-\frac{4}{3} \,.
\end{eqnarray}
One can see that  the first consistency condition in \eqref{eq:n1conditions} is satisfied. Further one finds \(\lambda=0\), which is also consistent with
(\ref{eq:Betherel}).
Equation (\ref{eq:spectralrelation}) gives now the relation between \(\Lambda\) and \(\alpha\)
\begin{equation}
 k^4\mathrm{sn}^4(\alpha)-2[(1+k^2)+\Omega^2]k^2\mathrm{sn}^2(\alpha)+(1+k^2+\Omega^2)^2=\Lambda,
\end{equation}
which can be solved as
\begin{equation}
 \mathrm{sn}(\alpha_{\pm}|k^2)=\sqrt{\frac{1+k^2\mp\sqrt{\Lambda}+\Omega^2}{k^2}},
\end{equation}
which agrees with the result (\ref{eq:solutionLame4}) directly obtained using the square property.

\section{Landau-Lifshitz  $SU(2)$ folded string analysis:  details}
\subsection{Spectral domain}
\label{sec:spectraldomain}

Useful properties of  $ \Omega_{\pm}$ defined in \eqref{eq:spectralLL}  are
\begin{eqnarray}\label{eq:omegadual}
 \Omega_{\pm}(-\alpha,k)&=&-\Omega_{\mp}(\alpha,k),\nonumber\\
 \Omega_{\pm}(\alpha+2i\mathbb{K}',k)&=&-\Omega_{\mp}(\alpha,k),\nonumber\\
 \Omega_{\pm}(\alpha+\mathbb{K}+i\mathbb{K}',k)&=&\mp\Omega_{\mp}(i\alpha,k'),\nonumber\\
 \Omega_{\pm}(\alpha+2\mathbb{K}+2i\mathbb{K}',k)&=&-\Omega_{\pm}(\alpha,k)~.
\end{eqnarray}
It  is then easy to see that  \(\Omega_{\pm}(\alpha)\) is a doubly periodic function
\begin{equation}
 \Omega_{\pm}(\alpha+4\mathbb{K},k)=\Omega_{\pm}(\alpha,k),\qquad\qquad \Omega_{\pm}(\alpha+4i\mathbb{K}',k)=\Omega_{\pm}(\alpha,k)
\end{equation}
Important special values are
\begin{eqnarray}
&& \Omega_+^2(\mathbb{K},k)=0,\qquad\qquad \Omega_+^2(i\mathbb{K}',k)=1,\nonumber\\
&&  \Omega_+^2\left(i\mathbb{K}'+i\mathrm{cn}^{-1}\left(\frac{k^2}{k'^2}\,,\,k'\right),k\right)=
 \Omega_+^2\left(i\mathbb{K}'-i\mathrm{cn}^{-1}\left(\frac{k^2}{k'^2}\,,\,k'\right)\right)=4k^2k'^2.
\end{eqnarray}
For the physical spectrum only those values of the complex parameter \(\alpha=u+iv\) corresponding to a real \(\Omega^2\) 
are of interest. We decompose
\begin{equation}
 \Omega_+(u+iv)=\mathrm{Re}(\Omega_+)(u,v)+i\,\mathrm{Im}(\Omega_+)(u,v)
\end{equation}
with
\begin{eqnarray}
 \!\!\!\!\mathrm{Re}(\Omega_+)(u,v)&=&\frac{2k\mathrm{dn}(u,k)(\mathrm{dn}(v,k')-k\mathrm{cn}(u,k)\mathrm{sn}(v,k'))}{(1-\mathrm{dn}^2(u,k)
 \mathrm{sn}^2(v,k'))^2}(\mathrm{cn}(u,k)\mathrm{cn}^2(v,k')
 +k\mathrm{sn}^2(u,k)\mathrm{sn}(v,k')\mathrm{dn}(v,k')),\nonumber\\\nonumber
 \!\!\!\!\mathrm{Im}(\Omega_+)(u,v)&=&\frac{2k\mathrm{sn}(u,k)\mathrm{cn}(v,k')(\mathrm{dn}(v,k')
 -k\mathrm{cn}(u,k)\mathrm{sn}(v,k'))}{(1-\mathrm{dn}^2(u,k)\mathrm{sn}^2(v,k'))^2}(k\mathrm{cn}(u,k)
 -\mathrm{dn}^2(u,k)\mathrm{sn}(v,k')\mathrm{dn}(v,k')).
\end{eqnarray}
In order to have \(\Omega_+^2\) real, we find the cases
\begin{itemize}
\item \(v=\mathbb{K'}\), then
\begin{equation}
 \Omega_+(u+i\mathbb{K}')=\frac{2\mathrm{dn}(u,k)}{\mathrm{sn}^2(u,k)}(1-\mathrm{cn}(u,k)).
\end{equation}
Setting \(u=2w\) gives
\begin{equation}
 \Omega_+(2w+i\mathbb{K}')=\mathrm{dc}^2(w,k)-k^2\mathrm{sn}^2(w,k).
\end{equation}
Varying \(w\) from \(0\) to \(\mathbb{K}\), then \(\Omega_+^2(2w+i\mathbb{K}')\) covers the interval \([1,\infty)\).
Therefore we can solve for \(w\) as
\begin{equation}
 \mathrm{sn}^2(w)=1-\frac{\Omega}{2k^2}+\frac{1}{2k^2}\sqrt{\Omega^2-4k^2k'^2} ,
\end{equation}
with
\begin{equation}
  0<\mathrm{sn}^2(w)<1\qquad\qquad\text{for}\qquad 1<\Omega<\infty.
\end{equation}
\item \(u=0\), then
\begin{equation}
 \Omega_+(iv)=\frac{2k}{\mathrm{cn}^2(v,k')}(\mathrm{dn}(v,k')-k\mathrm{sn}(v,k')).
\end{equation}
Varying \(v\) from \(\mathbb{K}'-\mathrm{cn}^{-1}(k^2/k'^2,k')\) to \(\mathbb{K}'\), then \(\Omega_+^2(iv)\) covers the interval \([4k^2k'^2,1]\).
Therefore we can solve for \(v\) as
\begin{equation}
 \mathrm{sn}^2(v,k')=1+\frac{4k^2}{\Omega^2}\left[-1+2k^2+\sqrt{\Omega^2-4k^2k'^2}\right],
\end{equation}
with
\begin{equation}
 \frac{k^2}{k'^2}<\mathrm{sn}^2(v,k')<1\qquad\qquad\text{for}\qquad 2kk'<\Omega<1\qquad\text{and}\qquad 
 0<k^2<\frac{1}{2}.
\end{equation}

\item \(k\mathrm{cn}(u,k)-\mathrm{dn}^2(u,k)\mathrm{sn}(v,k')\mathrm{dn}(v,k')=0\)

For \(0<u<\mathbb{K}\) this can be solved for \(v=v(u,k)\) as
\begin{equation}
 \mathrm{sn}^2(v,k')=\frac{1-\sqrt{1-4k^2k'^2\frac{\mathrm{cd}^2(u,k)}{\mathrm{dn}^2(u,k)}}}{2k'^2}\,,
\end{equation}
then 
\begin{equation}
 \alpha(u,k)=u+i\,\mathrm{sn}^{-1}\left[\frac{1}{\sqrt{2}k'}\sqrt{1-\sqrt{1-4k^2k'^2\frac{\mathrm{cd}^2(u,k)}{\mathrm{dn}^2(u,k)}}},k'\right]\,.
\end{equation}
After using some elliptic function identities one finds
\begin{equation}
 \Omega_+^2(\alpha(u,k),k)=4k^2k'^2\frac{\mathrm{cn}^2(u,k)}{\mathrm{dn}^4(u,k)}, \qquad\qquad 0<u<\mathbb{K}\,.
\end{equation}
Solving for \(u\) gives
\begin{equation}
 \mathrm{sn}^2(u,k)=\frac{1}{k^2}+\frac{2k'^2}{k^2}\frac{\sqrt{1-\Omega^2}-1}{\Omega^2},
\end{equation}
with
\begin{equation}
 0<\mathrm{sn}^2(u,k)<1,\qquad\qquad\text{for}\qquad 0<\Omega<2kk'\qquad\text{and}\qquad 0<k^2<\frac{1}{2}.
\end{equation}

\item  \(\mathrm{cn}(u,k)\mathrm{cn}^2(v,k')+k\mathrm{sn}^2(u,k)\mathrm{sn}(v,k')\mathrm{dn}(v,k')=0\)

For \(\mathbb{K}<u<2\mathbb{K}\) this can be solved for \(v\) as
\begin{eqnarray}
 \mathrm{sn}^2(v,k')&=&\frac{2\mathrm{cn}^2(u,k)}{2\mathrm{cn}^2(u,k)+k^2\mathrm{sn}^4(u,k)+k^2\mathrm{sn}^2(u)\sqrt{4\mathrm{cn}^2(u,k)
 +\mathrm{sn}^4(u,k)}}=\nonumber\\
 &=&\frac{2\mathrm{cn}^2(u,k)+k^2\mathrm{sn}^4(u,k)-k^2\mathrm{sn}^2(u,k)\sqrt{4\mathrm{cn}^2(u,k)+\mathrm{sn}^4(u,k)}}
 {2(\mathrm{cn}^2(u,k)+k^2k'^2\mathrm{sn}^4(u,k))}=\nonumber\\
&=&\frac{2\frac{\mathrm{cn}^2(u,k)}{\mathrm{sn}^4(u,k)}+k^2-k^2\sqrt{1+4\frac{\mathrm{cn}^2(u,k)}{\mathrm{sn}^4(u,k)}}}{2\left(k^2k'^2+
 \frac{\mathrm{cn}^2(u,k)}{\mathrm{sn}^4(u,k)}\right)}\,,
\end{eqnarray}
and then
\begin{equation}
 \alpha(u,k)=u+i\,\mathrm{sn}^{-1}\left[\sqrt{\frac{2\frac{\mathrm{cn}^2(u,k)}{\mathrm{sn}^4(u,k)}+k^2-k^2\sqrt{1+4\frac{\mathrm{cn}^2(u,k)}{\mathrm{sn}^4(u,k)}}}{2\left(k^2k'^2+
 \frac{\mathrm{cn}^2(u,k)}{\mathrm{sn}^4(u,k)}\right)}},k'\right]\,.
\end{equation}
After using some elliptic function identities one finds
\begin{equation}
 \Omega_+^2(\alpha(u),k)=-\frac{4\mathrm{cn}^2(u,k)}{\mathrm{sn}^4(u,k)},\qquad\qquad\text{for}\qquad \mathbb{K}<u<2\mathbb{K}\,,
\end{equation}
or
\begin{equation}
\Omega_+^2(\alpha(\tilde u+\mathbb{K}),k)=-4k'^2\frac{\mathrm{sn}^2(\tilde u,k)\mathrm{dn}^2(\tilde u,k)}{\mathrm{cn}^4(\tilde u,k)}\,.
\end{equation}
Solving for \(u\) gives
\begin{equation}
\mathrm{sn}^2(u,k)=\frac{2}{\Omega^2}(1-\sqrt{1-\Omega^2})\,,
\end{equation}
with
\begin{equation}
 0<\mathrm{sn}^2(u,k)<1\qquad\qquad\text{for}\qquad -\infty<\Omega^2<0\,.
\end{equation}
\end{itemize}
 


The expressions of $\alpha_i$'s in the different branches for $ \Omega^2$ read  as follows:
\begin{itemize}
\item For \(-\infty<\Omega^2<0\) as
\begin{eqnarray}\label{alphabranchnegag}
 \alpha_1(\Omega,k)&=&u(\Omega,k)-iv(\Omega,k),\nonumber\\
 \alpha_2(\Omega,k)&=&2\mathbb{K}-u(\Omega,k)+iv(\Omega,k),\nonumber\\
\alpha_3(\Omega,k)&=&2\mathbb{K}+u(\Omega,k)+iv(\Omega,k),\nonumber\\
 \alpha_4(\Omega,k)&=&2\mathbb{K}-u(\Omega,k)+2i\mathbb{K}'-iv(\Omega,k),
\end{eqnarray}
where
\begin{equation}\label{alphabranchnegaguv}
 \!\!\!\!\!\!
 \!\!\!\!\!\!
 u(\Omega,k)=\mathrm{sn}^{-1}\left[\sqrt{\frac{2}{\Omega^2}(1-\sqrt{1-\Omega^2})},\,k\right],
 \qquad
 v(\Omega,k)=\mathrm{sn}^{-1}\left[\sqrt{\frac{\Omega^2-2k^2+2k^2\sqrt{1-\Omega^2}}{\Omega^2-4k^2k'^2}},\,k'\right]\,.
\end{equation}
\item For \(0<\Omega^2<4k^2k'^2\) as
 \begin{eqnarray}
 \alpha_1(\Omega,k)&=&2\mathbb{K}-u_2(\Omega,k)-iv_2(\Omega,k),\nonumber\\
 \alpha_2(\Omega,k)&=&u_2(\Omega,k)+iv_2(\Omega,k),\nonumber\\
 \alpha_3(\Omega,k)&=&2\mathbb{K}+u_2(\Omega,k)-iv_2(\Omega,k),\nonumber\\\label{alphabranchpos1}
 \alpha_4(\Omega,k)&=&2\mathbb{K}-u_2(\Omega,k)+2i\mathbb{K}'+iv_2(\Omega,k),
\end{eqnarray}
where
\begin{equation}
 u_2(\Omega,k)=\mathrm{sn}^{-1}\left[\frac{1}{k}\sqrt{1-2k'^2\frac{1-\sqrt{1-\Omega^2}}{\Omega^2}}\right],\qquad\qquad 
 v_2(\Omega,k)=\mathrm{sn}^{-1}\left[\frac{1}{\sqrt{2}k'}\sqrt{1-\sqrt{1-\Omega^2}},k'\right]\,.
\end{equation}
\item For \(4k^2k'^2<\Omega^2<\infty\) as
\begin{eqnarray}
 \alpha_3(\Omega,k)&=&2\mathbb{K}-i\mathbb{K}'+2i\alpha_0(\Omega,k'),\nonumber\\
 \alpha_4(\Omega,k)&=&2\mathbb{K}+3i\mathbb{K}'-2i\alpha_0(\Omega,k')\,.
\end{eqnarray}

\item For \(4k^2k'^2<\Omega^2<1\)
\begin{eqnarray}
 \alpha_1(\Omega,k)&=&2\mathbb{K}-i\,\mathrm{sn}^{-1}\left[\sqrt{1-\frac{4k^2}{\Omega^2}\left(1-2k^2-\sqrt{\Omega^2-4k^2k'^2}\right)},k'\right],\nonumber\\
 \alpha_2(\Omega,k)&=&i\,\mathrm{sn}^{-1}\left[\sqrt{1-\frac{4k^2}{\Omega^2}\left(1-2k^2-\sqrt{\Omega^2-4k^2k'^2}\right)},k'\right].
\end{eqnarray}
\item For \(1<\Omega^2<\infty\) as
\begin{eqnarray}
 \alpha_1(\Omega,k)&=&2\mathbb{K}-i\mathbb{K}'-2\alpha_0(\Omega,k)\nonumber\\
 \alpha_2(\Omega,k)&=&i\mathbb{K}'+2\alpha_0(\Omega,k), 
\end{eqnarray}
where
\begin{equation}\label{alphabranchpos3}
\alpha_0(\Omega,k)=\mathrm{sn}^{-1}\left[\sqrt{1-\frac{\Omega}{2k^2}+\frac{1}{2k^2}\sqrt{\Omega^2-4k^2k'^2}},\,k\right]\,.
\end{equation}
\end{itemize}

\subsection{A duality property of the LL fourth order differential operator }
\label{app:duality}

Defining \(z=ix\)  we can rewrite (\ref{LL_first}) as
\begin{eqnarray}
 \mathcal{O}^{(4)}(z,k')f_{1,2}(-iz-\mathbb{K}+i\mathbb{K}',\alpha,k)&=&\Omega_-^2(\alpha,k)f_{1,2}(-iz-\mathbb{K}+i\mathbb{K}',\alpha,k),
 \nonumber\\
\mathcal{O}^{(4)}(z,k')f_{3,4}(-iz-\mathbb{K}+i\mathbb{K}',\alpha,k)&=&\Omega_+^2(\alpha,k)f_{3,4}(-iz-\mathbb{K}+i\mathbb{K}',\alpha,k).
\end{eqnarray}
Interchanging the role of \(k\) and \(k'\) and using (\ref{eq:omegadual}) we get
\begin{equation}
 \mathcal{O}^{(4)}(x,k)f_{3,4}(-ix-\mathbb{K}'+i\mathbb{K},i\alpha-i\mathbb{K}+\mathbb{K}',k')=\Omega_-^2(\alpha,k)
f_{3,4}(-ix-\mathbb{K}'+i\mathbb{K},i\alpha-i\mathbb{K}+\mathbb{K}',k')\,.
\end{equation}
Using elliptic function identities one can show that
\begin{equation}
 f_3(-ix-\mathbb{K}'+i\mathbb{K},i\alpha-i\mathbb{K}+\mathbb{K}',k')=c(\alpha,k)f_2(x,\alpha,k),
\end{equation}
with a \(x\)-independent constant
\begin{equation}
 c(\alpha,k)=\exp\left[\frac{\pi}{4\mathbb{K}\mathbb{K}'}(\mathbb{K}-\alpha)^2-\frac{i\pi\alpha}{2\mathbb{K}}+
 (\mathbb{K}'-i\mathbb{K})(-iZ(\alpha,k)+k\mathrm{cn}(\alpha,k))\right]\,.
\end{equation}
The duality implies that an eigenfunction for $\Omega_-^2$ and $0<k^2 < 1/2$
becomes an eigenfunction for $\Omega_+^2$ and $1/2 < k^2 < 1$. An analogous relation holds for $f_1$ and $f_4$.

\subsection{Finite-gap structure: a microscopical spectral curve}
\label{app:spectral}

To uncover the finite-gap structure of the semi-classical fluctuation spectral problem governed by the fourth 
order differential operator  \eqref{LL_first}
one starts by evaluating the differential of the quasi-momentum function (\ref{quasimomentaLL}),
entering the eigenfunctions of the LL operator (\ref{LL_first}), in two steps according to
\begin{equation}
 \frac{\mathrm{d}p}{\mathrm{d}(\Omega^2)}=\frac{\mathrm{d}p}{\mathrm{d}\alpha}\frac{\mathrm{d}\alpha}{\mathrm{d}(\Omega^2)}.
\end{equation}
As an example, we choose  \(-\infty<\Omega^2<0\) and plug (\ref{alphabranchnegag})-(\ref{alphabranchnegaguv}) 
into (\ref{quasimomentaLL}). Choosing the two linearly independent quasi-momenta as in (\ref{p1andp2}) 
we get 
\begin{eqnarray}
 \frac{\mathrm{d}p_1}{\mathrm{d}(\Omega^2)}&=&i\frac{-k'^2-\frac{i}{2}\sqrt{4k^2k'^2-\Omega^2}+\frac{\mathbb{E}}{
 \mathbb{K}}}{2\sqrt{-\Omega^2}\sqrt{-1+2k^2-i\sqrt{4k^2k'^2-\Omega^2}}\sqrt{4k^2-4k^4-\Omega^2}}\\
 \frac{\mathrm{d}p_2}{\mathrm{d}(\Omega^2)}&=&i\frac{k'^2-\frac{i}{2}\sqrt{4k^2k'^2-\Omega^2}-\frac{\mathbb{E}}{
 \mathbb{K}}}{ 
 2\sqrt{-\Omega^2}\sqrt{-1+2k^2+i\sqrt{4k^2k'^2-\Omega^2}}\sqrt{4k^2-4k^4-\Omega^2}}~,
\end{eqnarray}
 and similarly for the three branches covering the positive-frequency range (see Appendix 
\ref{sec:spectraldomain}).
Focussing on the first of these formulas, we introduce a new spectral parameter 
\begin{equation}
 z=\frac{1}{2}\sqrt{\Omega^2-4k^2(1-k^2)}
\end{equation}
which results in the following differential of the  quasi-momentum $p_1$ 
\begin{eqnarray}\label{finitegap}
 \frac{\mathrm{d}p_1}{\mathrm{d}z} = \frac{z+z_0}{\sqrt{2(z-z_1)(z-z_2)(z-z_3)}}
\end{eqnarray}
with
\begin{equation} 
 z_0=1-k^2-\frac{\mathbb{E}}{\mathbb{K}},\qquad\qquad z_1=-ikk',\qquad z_2=ikk',\qquad 
 z_3=k^2-\frac{1}{2}.
\end{equation}
The set of points described by the elliptic curve $y^2=(z-z_1)(z-z_2)(z-z_3)$ on 
the complex $z$ plane defines what one could call a ``microscopical'' spectral curve for the LL string action, in the sense that 
it encodes the dynamics of the one-loop fluctuations above the classical, ``macroscopical'' spectral curve 
emerging in the finite-gap picture of~\cite{Kazakov:2004qf}. The differential \eqref{finitegap} clearly uncovers the one-gap
structure of the corresponding spectral curve and justifies to consider the corresponding differential equations as 
fourth order analogs of the  second order, finite-gap, Lam\'e operators 
of~\cite{Beccaria:2010ry}.



\subsection{The short string  expansion}
\label{app:shortstring}
 
In this appendix we spell out the expansion of $p_1$ and $p_2$ as series of $k$ in the physical branch $\Omega^2<0$. 
The starting point is the expression of the quasi-momentum function (\ref{quasimomentaLL}), 
evaluated on the four functions $\alpha_i$ given in (\ref{alphabranchnegag}). 
As recalled in Section \ref{sec:4thorder}, the corresponding values of the momenta are not all linearly independent, and  we choose 
\begin{flalign}
p_{1} & = i Z\left(\alpha_{2},k\right) + k \mathrm{cn}\left(\alpha_{2},k\right)-\frac{\pi}{2 \KK}\,, \nn\\
\label{p1andp2}
p_{2} & = i Z\left(\alpha_{3},k\right) + k \mathrm{cn}\left(\alpha_{3},k\right)-\frac{\pi}{2 \KK}.
\end{flalign}
In the following we will provide the details on the expansion of the former, since the analysis of the latter proceeds in the same fashion, modulo some minus 
signs.
\\
Using the Jacobi Zeta function the addition  formula for complex argument leads to
\begin{eqnarray}
p_{1}&= 
&-iZ\left(u,k\right)+Z\left(v,k'\right)-k\mathrm{cn}\left(u-iv,k\right)+\frac{v\pi}{2\KK\KK^{'}}-\frac{\pi}{2\KK}  \\\nonumber
&& -i\frac{k^{2}\mathrm{sn}\left(u,k\right)\mathrm{cn}\left(u,k\right)\mathrm{dn}\left(u,k\right)\mathrm{sn}^{2}\left(v,k'\right)}{1-\mathrm{sn}^{2}\left(v,k'\right)\mathrm{dn}^{2}\left(u,k\right)}
-\frac{\mathrm{dn}^{2}\left(u,k\right)\mathrm{cn}\left(v,k'\right)\mathrm{sn}\left(v,k'\right)\mathrm{dn}\left(v,k'\right)}{1-\mathrm{sn}^{2}\left(v,k'\right)\mathrm{dn}^{2}\left(u,k\right)}.
\end{eqnarray}
The functions $u\left(\Omega,k\right)$ and $v\left(\Omega,k\right)$, parametrizing the real and imaginary part of $\alpha_2$ respectively, were
presented in (\ref{alphabranchnegaguv}). The $k\sim0$ expansion of some terms above 
can be treated 
expanding  their derivatives with respect to $\sqrt{\Omega^2}$ and 
integrating back at the end. The derivative 
of the Jacobi Zeta function with argument $0<x<\KK$, the function $x$ being $u\left(\Omega,k\right)$ or $v\left(\Omega,k\right)$, is conveniently 
expressed as
\begin{flalign}
\frac{\partial Z\left(x,k\right)}{\partial\left(\sqrt{\Omega^2}\right)}=\frac{1}{\sqrt{1-\mathrm{sn}^2\left(x,k\right)}\sqrt{1-k^2\mathrm{sn}^2\left(x,k\right)}} 
\frac{\partial\mathrm{sn}\left(x,k\right)}{\partial\left(\sqrt{\Omega^2}\right)}\left[1-k^2\mathrm{sn}^2\left(x,k\right)-\frac{\EE}{\KK}\right].
\end{flalign}
Up to fourth order, the expansion for the two independent momenta reads
\begin{flalign}
p_{1} & = -i\sqrt{-1-i\sqrt{\Omega^2}}+\frac{\left(2-i\sqrt{\Omega^2}\right)\sqrt{1+i\sqrt{\Omega^2}}}{8\Omega^{2}}k^{4}+\mathcal{O}\left(k^{6}\right)\\
p_{2} & 
=+i\sqrt{-1+i\sqrt{\Omega^2}}+\frac{\left(2+i\sqrt{\Omega^2}\right)\sqrt{1-i\sqrt{\Omega^2}}}{8\Omega^{2}}k^{4}+\mathcal{O}\left(k^{6}\right)\,,
\end{flalign}
to which corresponds the expansion in the regularized effective action
(\ref{Gamma_reg}). 
The latter can be written as
\be
\Gamma^{(1)}_{reg}= \sum_{i=0}^{\infty}\Gamma_{i, reg}^{(1)} k^{2i}\,,
\ee
where the first term is vanishing by construction.  
The quasi-momenta in  \eqref{Gamma_reg} are  computed in the physical region $\Omega^2<0$, 
which corresponds to an Euclidean partition function. However, we find convenient to perform our 
integrals by analytically-continuing all the expressions to $\Omega^2\to -\Omega^2$~\footnote{In this 
way, see \eqref{poles}, the poles of the integrand are on the imaginary axis. Alternatively, 
one could not perform the analytic continuation and  treat the poles appearing on the real axis with  an $i\epsilon$ prescription. }.
This results in the following expressions for the first few terms
\be
\Gamma_{0, reg}^{(1)} &=& 0\,,\\
\label{Gamma_reg_int_1}
\frac{\Gamma^{(1)}_{1, reg} }{\cal T} &=&
\frac{\pi}{8}\int {d\Omega\over 2\pi} \left[\frac{\sqrt{-1-i\sqrt{\Omega^{2}}}}{\tanh\left(\pi\sqrt{-1-i\sqrt{\Omega^{2}}}\right)}+\frac{\sqrt{-1+i\sqrt{\Omega^{2}}}}{\tanh\left(\pi\sqrt{-1+i\sqrt{\Omega^{2}}}\right)}\right] \,,\\ 
\label{Gamma_reg_int_2}
\frac{\Gamma^{(1)}_{2, reg}}{\cal T} & = & \frac{\pi}{32}\int\frac{d\Omega}{2\pi}\left[-\pi+\frac{\pi\left(1+i\sqrt{\Omega^{2}}\right)}{2\tanh^{2}\left(\pi\sqrt{-1-i\sqrt{\Omega^{2}}}\right)}+\frac{\pi\left(1-i\sqrt{\Omega^{2}}\right)}{2\tanh^{2}\left(\pi\sqrt{-1+i\sqrt{\Omega^{2}}}\right)}\right.\nonumber\\
 &  & \hphantom{\frac{\pi}{32}\int\frac{d\Omega}{2\pi}}+\frac{\left(-16+8i\sqrt{\Omega^{2}}+17\Omega^{2}\right)\sqrt{-1-i\sqrt{\Omega^{2}}}}{4\Omega^{2}\tanh\left(\pi\sqrt{-1-i\sqrt{\Omega^{2}}}\right)}\nonumber\\
 &  & 
 \hphantom{\frac{\pi}{32}\int\frac{d\Omega}{2\pi}}\left.+\frac{\left(-16-8i\sqrt{\Omega^{2}}+17\Omega^{2}\right)\sqrt{-1+i\sqrt{\Omega^{2}}}}{4\Omega^{2}\tanh\left(\pi\sqrt{-1+i\sqrt{\Omega^{2}}}\right)}\right],
\ee
where we notice that there is no obstruction to go to higher terms. 
The integrals above are divergent, which is due to the absence in the LL action of fermionic and 
some bosonic modes which are crucial for UV finiteness.
A first form
of regularization is realized  embedding our real integrals (\ref{Gamma_reg_int_1})-(\ref{Gamma_reg_int_2}) 
in the complex plane, in order to exploit Cauchy's residue theorem.
The integrands turn out to be meromorphic functions on $\mathbb{C}\setminus \{0\}$, featuring poles on the imaginary axis 
at
\be\label{poles}
\Omega^{\pm}_{n}\equiv\pm i \left|n^2-1\right|, \qquad n=2,3,... .
\ee
By closing the contour of integration on the anti-clockwise upper-half (clockwise lower-half) plane, the finite part is then conventionally defined to be the $\zeta$-regularized sum of 
the residues, dropping possibly divergent contributions from the arc wrapping the poles $\Omega^{+}_{n}$ (resp. $\Omega^{-}_{n}$). 
This prescription
brings to the finite answers
\begin{flalign}
\frac{\Gamma_{1, reg}^{(1)} }{\cal T}& = 2\pi i \sum_{n=2}^{\infty} \frac{i n^2}{8\pi}= 
\frac{1}{4}\\
\frac{\Gamma_{2, reg}^{(1)}}{\cal T} & = 2\pi i \sum_{n=2}^{\infty} \frac{i n^2\left(35-30n^2+11n^4\right)}{128\pi \left(n^2-1\right)^2} =
\left(\frac{1}{16}-\frac{\pi^2}{48}\right),
\end{flalign}
which finally lead to the expected one-loop energy (\ref{one_loop_energy}).
\\
Alternatively, we can cut off the frequency domain $\epsilon<\left|\Omega\right|<L$ 
and safely work the real integrals out by employing the infinite sum representation for $\coth$ and $\coth^2$
\begin{flalign}
\coth\pi x & = \frac{1}{\pi x}+\frac{2}{\pi}\sum_{n=1}^{\infty} \frac{x}{n^2+x^2}\\
\coth^2\pi x & = 1+\frac{1}{\pi^2 x^2}-\frac{2}{\pi^2}\sum_{n=1}^{\infty} \frac{n^2-x^2}{\left(n^2+x^2\right)^2}.
\end{flalign}
The trick consists in performing the integrations first
\begin{flalign}
  \Gamma_{1, reg}^{(1)} & =\lim_{L\to\infty}\left[\frac{3L}{4\pi}+\sum_{n=2}^{\infty}\left(-\frac{n^{2}}{4}+\frac{L}{2\pi}\right)\right]~,\\
  \Gamma_{2, reg}^{(1)} & =\lim_{L\to\infty}\lim_{\epsilon\to0}\left\{ \frac{45L}{64\pi}-\frac{3}{8\pi\epsilon}+\sum_{n=2}^{\infty}\left[-\frac{11n^{2}}{64}+\frac{15L}{32\pi}+\frac{1}{8}\right.\right.\nonumber\\
  & \hphantom{=\lim_{L\to+\infty}\lim_{\epsilon\to0^{+}}}\left.\left.-\frac{1}{16\left(n+1\right)^{2}}-\frac{1}{16\left(n-1\right)^{2}}+\frac{1}{2\pi\left(n^{2}-1\right)\epsilon}\right]\right\},
\end{flalign}
followed by the $\zeta$-regularized sums. Upon $\zeta$-regularization, the $\epsilon$- and 1/L-coefficients are finite
and drop out in the limit. On the other hand, UV- and IR-divergencies happen to cancel out 
and the cut-off regularization reproduces the same one-loop energy contribution  
(\ref{one_loop_energy}) of the residue prescription.


\section{Folded string in full sigma-model:  details}
\label{app:foldedfulldetails}

\subsection{Fluctuation Lagrangian in static gauge}
\label{app:foldedfullstatic}

A fluctuation Lagrangian derived from Pohlmeyer reduction
 that agrees with the Nambu action in static gauge (in which fluctuations of 
$t$ and $\rho$ are set to zero)    result is~\foot{We thank I.
Iwashita, R. Roiban and A. A. Tseytlin for this information. There are other more complicated forms of the  fluctuation action
that should be  related by field redefinitions.}  
\ba
&&L= \p_a x \p^a x - \mu^2_x  x^2  + \p_a y \p^a y - \mu^2_y  y^2
 +  2 q   \  y \p_ 0 x  \ , \la{ki}\\
 && \ \ \ \ \  \ \ \ 
 q= { 2 \nu \bar{w} \k \ov  \r'^2 + \n^2}   \la{kqi}\\
&& \la{ji} 
\mu^2_x= 2 \r'^2 + \n^2  +  {\k^2 \bar{w}^2 ( 2 \r'^2 - \n^2) \ov (  \r'^2 + \n^2)^2 }  \ , \ \ \ \ \ 
\mu^2_y = \nu^2  \Big[ -1 + { 2 ( \bar{w}^2 + \k^2) \ov   \r'^2 + \n^2} 
-   {  3 \bar{w}^2 \k^2  \ov   (\r'^2 + \n^2)^2 } \Big]  
\ea 
Here $x$ and $y$  are two physical fluctuations in $AdS_3$ sector.
When $\nu \to 0$ we get  one massless mode and a mode with 
$\mu^2 = 2 \r'^2 + {2\k^2 \bar{w}^2  \ov   \r'^2 }$ as  expected.

All other fluctuations have same mass as in the conformal gauge  
discussed in \cite{Frolov:2002av}, e.g., 
 $\beta_u$ ($u=1,2$) -- the  fluctuations in $AdS_5$  that are 
transverse to $AdS_3$ --  have mass $\mu_\b^2 = 2 \r'^2 + \nu^2$.

The corresponding equations of motion  are ($ x \to e^{i \omega \tau} \td x(\s) , \ \ 
y \to e^{i \omega \tau} \td y(\s)$) \foot{Here we did not yet switch
 to Euclidean time $\tau \to i \tau$, i.e.    $\o \to i \o$.}  
\be 
(\p_\s^2  + \omega^2 - \m_x^2) \td x  -i q \,\omega\,  \td y =0 \ , \ \ \ \ \ \ \ \ 
(\p_\s^2  + \omega^2 - \m_y^2) \td y  + i   q\, \omega\,  \td x =0 \ . 
\la{py}
\ee 
Here $q,\mu_x,\mu_y$ depend on $\s$. Solving the first equation  for $\td y$ 
and substituting into the second equation, we get 
 fourth order equation 
\be \la{ko}
\Big[ (\p_\s^2  + \omega^2 - \m_x^2) [  {1 \ov  q}  (\p_\s^2  + \omega^2 - \m_y^2)]  - q  \omega^2 \Big] \td y =0 
\ee
or explicitly
\be 
\Big[ (\p_\s^2  + \omega^2 - \m_x^2)  [(\r'^2 +\nu^2)   (\p_\s^2  + \omega^2 - \m_y^2)]   
- { 4 \nu^2 \k^2 \bar{w}^2 \ov
\r'^2 + \nu^2 }    \omega^2 \Big] \td y =0  \la{hh}\,. 
\ee
Redefining
\be
\tilde y=\frac{1}{\sqrt{\r'^2+\nu^2}}\,\tilde\r
\ee
one recovers the fourth-order differential equation
\be
\nn
\Big[\p^4_\s +2(\bar{w}^2+\k^2-2\nu^2+ \omega^2-4\r'^2) \ \p^2_\s 
 -   8\r' \r''\  \p_\s +  \k^4+(\bar w^2-\omega^2)^2 - 2\k^2(\omega^2+\bar w^2)\Big]\tilde\r=0\,,\\
\ee
which coincides with the one \eqref{eq:opnunonzero2} obtained in conformal gauge, once one uses the classical equation of motion for $\rho$ \eqref{eom}, and performs a Wick rotation $\Omega^2\to -\Omega^2$.

\bigskip

In the long string limit the operator \eqref{ko} agrees with the results first found in \cite{Frolov:2006qe}. 
This limit  (also known as large spin regime) 
corresponds to
\be\label{largespin} 
\r= \mu \s \ , \ \ \ \  \ \bar w=\k = \sqrt{ \mu^2 + \nu^2}  \ , \ \ \ \   \ \mu= { 1 \ov \pi} \log S \gg 1 \  , 
\ee
and the mass operators \eqref{kqi}-\eqref{ji} become
\be 
\mu_x^2 \to  4 \mu^2 \ , \ \ \ \ \ \ \mu_y^2 \to 0 \ , \ \ \ \ \ 
q\to 2 \nu  \ , \ee
and thus \eqref{ko}  (with $\p_\s \to i n$)  gives the following characteristic polynomial 
\be 
(\omega^2-n^2 - 4\mu^2) (\omega^2-n^2) - 4 \nu^2 \omega^2 =0  \ , \la{rr} \ee
with 
\be\label{goodfreq} 
\omega^2 = n^2  + 2 (\mu^2 + \nu^2) \pm 2 \sqrt{ n^2 \nu^2 + (\mu^2 + \nu^2)^2}  \ , 
\ee
which agrees with the expression in \cite{Frolov:2006qe}  
where $\omega$ and $n$ are rescaled by $\k= \sqrt{ \mu^2 + \nu^2}$, i.e. ($p= {n \ov \k} $)
\be 
\bar \Omega^2 = p^2  + 2  \pm 2 \sqrt{ p^2 u^2 + 1}  \ , \ \ \ \ \ \ \    u \equiv  { \nu \ov \k}  \ .  \ee
Taking into account the other masses for the remaining fluctuations in this long string limit, that is
\begin{enumerate}[-]
\item
two transverse fluctuations in  $AdS_5$:  $\mu_\b^2 = 2 \r'^2 + \nu^2\to  2 \mu^2 + \nu^2 = \k^2 ( 2 - u^2) $;
\item
four  fluctuations in  $S^5$:  $\mu_{sph}^2 = \nu^2 = \k^2   u^2 $; 
\item 
eight fermionic fluctuations $ \mu_\psi^2 = \r'^2 + \nu^2 = \k^2 $, 
\end{enumerate}
the corresponding characteristic frequencies \eqref{goodfreq} produce the well-known one-loop expression 
for the energy in this scaling limit \cite{Frolov:2006qe}.  

\subsection{Spectral domain and four linear independent solutions}
\label{sec:spectraldomain_folded}

We can repeat the analysis of the previous section \ref{sec:spectraldomain} for the folded string. 
Here calculations are carried out in Minkowski signature. 
Again, from the consistency equations we obtain the relations
\be
&& \lambda=\pm\sqrt{k^2\mathrm{sn}^2(\alpha)-\bar\Omega^2}\,,
\nn \\ 
\label{eq:spectralpara1}
&&  \bar\Omega^2_{\pm}(\alpha)=-(\kappa^2-\nu^2)\mathrm{sn}^2(\alpha)\left(\frac{\bar w k\,\mathrm{cn}(\alpha)\pm\kappa\,\mathrm{dn}(\alpha)}{(\kappa^2-\nu^2)\mathrm{sn}^2(\alpha)+\nu^2}\right)^2\,.
\ee
As before it is useful to work out duality relations for the eigenvalues, that is
\begin{eqnarray}
 \bar\Omega_{\pm}^2(\alpha+2\mathbb{K})=\bar\Omega_{\mp}^2(\alpha)\,, \qquad 
 \bar\Omega_{\pm}^2(\alpha+2i\mathbb{K}')=\bar\Omega_{\pm}^2(\alpha)\,,
\end{eqnarray}
which allows us to work only with  one kind of rescaled frequency, that is \(\bar\Omega(\alpha)\equiv\bar\Omega_+(\alpha)\).
In order to further manipulate the expression for $\Omega$ and find the corresponding ``physical'' spectral curve, it is advantageous to introduce 
the function
\begin{equation}
 a^2(\alpha)=(\kappa^2-\nu^2)\mathrm{sn}^2(\alpha)+\nu^2\,,
\end{equation}
with the property
\begin{equation}
\frac{\mathrm{d}}{\mathrm{d}\alpha}a^2(\alpha)=\frac{2}{\sqrt{\bar w^2-\nu^2}}\sqrt{a^2(\alpha)-\nu^2}\sqrt{\kappa^2-a^2(\alpha)}
 \sqrt{\bar w^2-a^2(\alpha)}\,.
\end{equation}
Thus, one can rewrite (\ref{eq:spectralpara1}) as follows
\begin{equation}
 \bar\Omega_{\pm}^2(\alpha)=-\frac{1}{\bar w^2-\nu^2}\frac{a^2(\alpha)-\nu^2}{a^4(\alpha)}
 \left(\bar w\sqrt{\kappa^2-a^2(\alpha)}\pm\kappa\sqrt{\bar w^2-a^2(\alpha)}\right)^2\,,
\end{equation}
and it easily follows now
\begin{eqnarray}
 \frac{\partial\bar\Omega^2}{\partial\alpha}&=&
2\bar\Omega^2(\alpha)\frac{1}{(\kappa^2-\nu^2)\mathrm{sn}^2(\alpha)+\nu^2}\left(\nu^2\frac{\mathrm{cn}(\alpha)\mathrm{dn}(\alpha)}{\mathrm{sn}(\alpha)}+k\, \kappa\, \bar w\, \mathrm{sn}(\alpha)\right)\,,
\end{eqnarray}
\begin{eqnarray}\label{eq:LambdaNu}
 \lambda(\alpha)&=&
 \frac{k\,\mathrm{sn}(\alpha)}{(\kappa^2-\nu^2)\mathrm{sn}^2(\alpha)+\nu^2}\left(\kappa\bar w\, \pm\sqrt{\kappa^2-\nu^2}\sqrt{\bar w^2-\nu^2}\,
 \mathrm{cn}(\alpha)\mathrm{dn}(\alpha)\right)\,.
\end{eqnarray}

\begin{figure}[h]
 \includegraphics[scale=0.45]{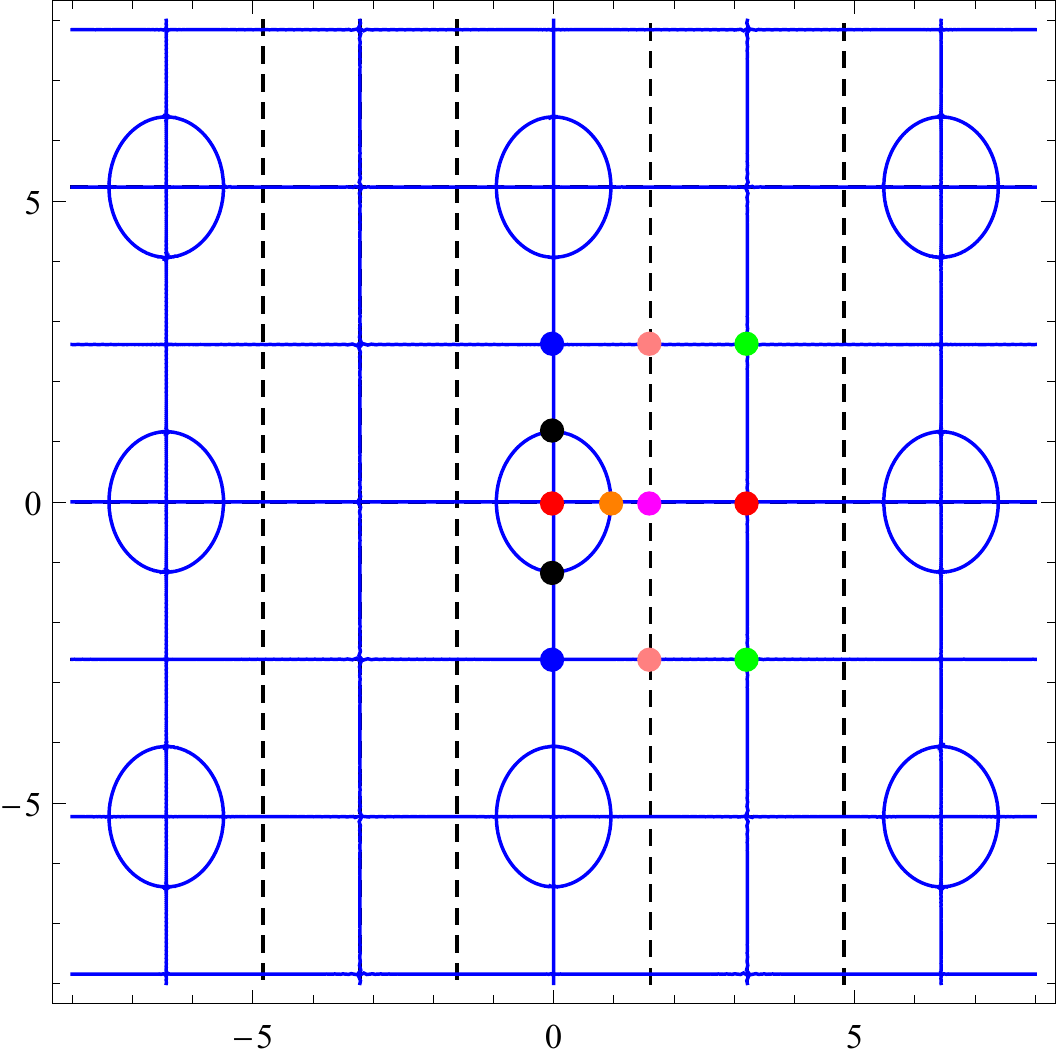}
 \includegraphics[scale=0.45]{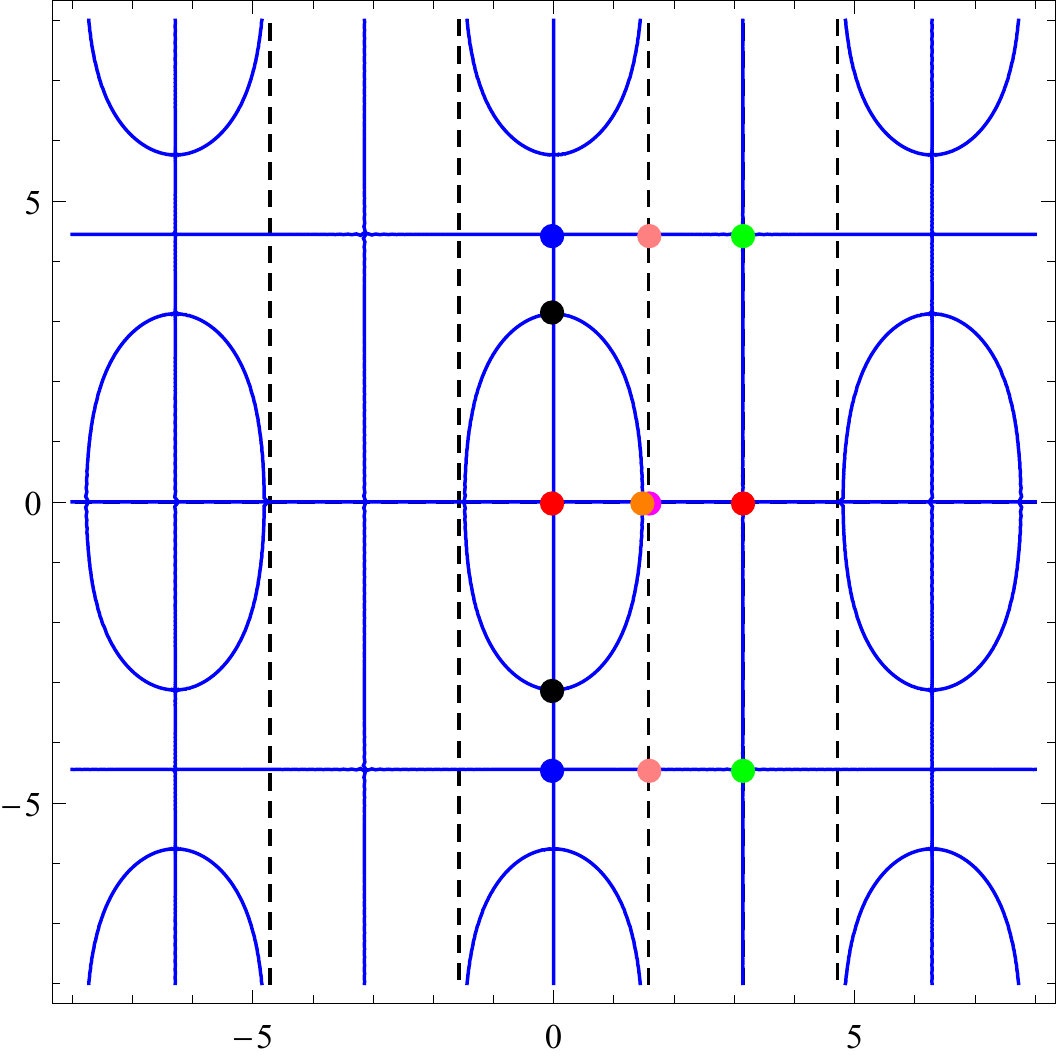}
 \includegraphics[scale=0.45]{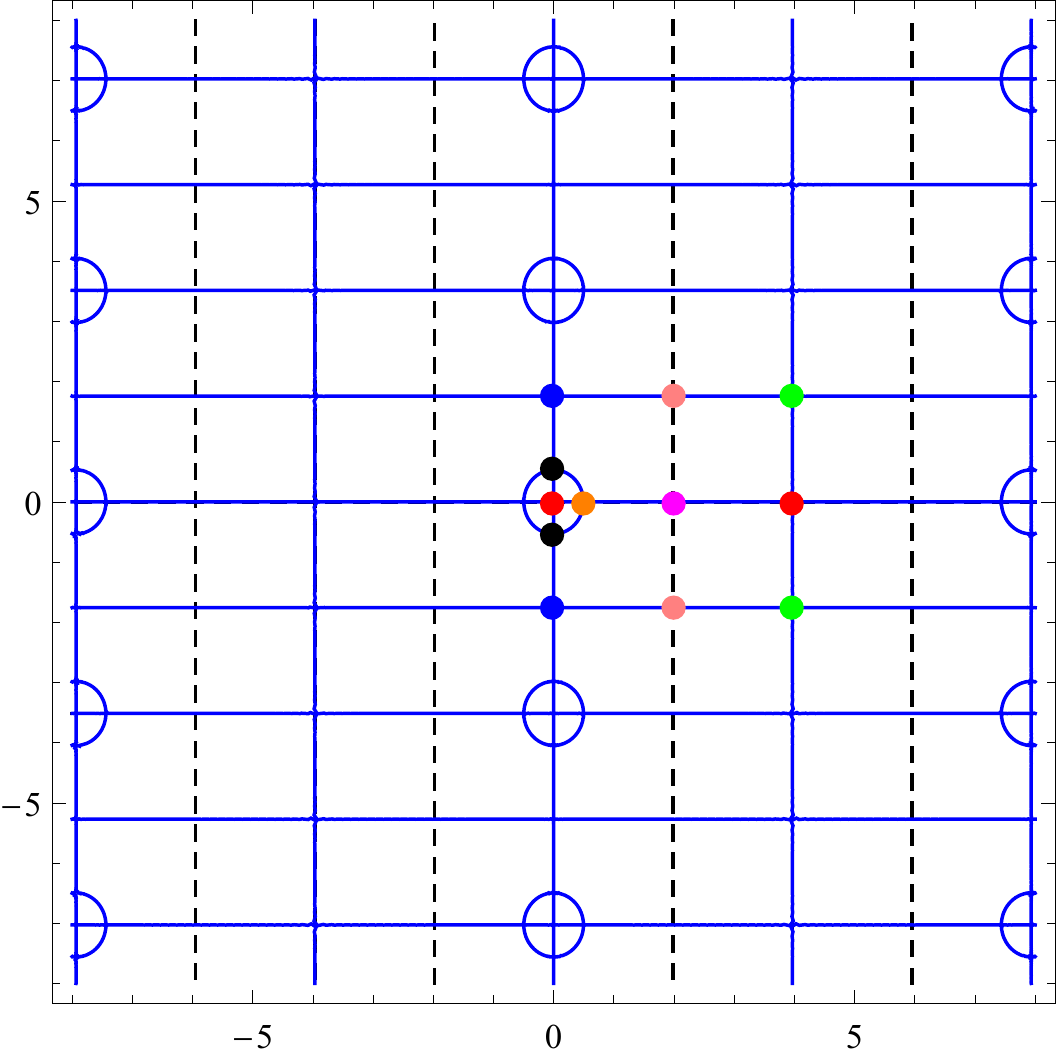}
 \caption{In the complex \(\alpha\) plane one can plot the lines where \(\bar\Omega^2(\alpha)\) is real. The parameters are chosen on the left as \(\kappa=3\), \(\omega=6\),
  \(\nu=2.5\) (\(k\sim 0.1\)), in the middle as \(\kappa=3\), \(\omega=6\), \(\nu=2.99\) (\(k\sim 0.002\)) and on the right as \(\kappa=5\), \(\omega=6\), \(\nu=2.5\) (\(k\sim 0.63\)). The special points are marked with colors as follows 
 \(\bar\Omega^2=-(\frac{\kappa^2}{\nu^2}-1)\) (orange), 
 \(\bar\Omega^2=-\textstyle\frac{(\kappa^2-\nu^2)(\omega^2-\kappa^2)}{\kappa^2(\omega^2-\nu^2)}\) (magenta),  \(\bar\Omega^2=0\) (red), 
 \(\bar\Omega^2=\frac{(\omega-\kappa)^2}{\omega^2-\nu^2}\) (green), \(\bar\Omega^2=1-\frac{\kappa^2}{\omega^2}\) (pink), 
\(\bar\Omega^2=\frac{(\omega+\kappa)^2}{\omega^2-\nu^2}\) (blue).}
\label{fig:FSPlane1}
\end{figure}

We are interested in all values of the complex parameter \(\alpha=u+iv\) that correspond to real values of \(\bar\Omega^2\).
The condition \(\mathrm{Im}(\bar\Omega^2)(u,v)=0\) results in the cases \(u=0,\,\, i\mathbb{K}'\), \(v=0,\,\,2\mathbb{K}\) (modulo periodicity of 
 \(\bar\Omega^2(\alpha)\)), which
gives the straight lines in Fig \ref{fig:FSPlane1}. 
However, this does not exhaust all the possibilities: There are still orbits in the $\alpha$ complex plane where $\bar\Omega^2$ is real, which correspond to the ellipse-like lines in Fig. \ref{fig:FSPlane1}. In order to find a parametrization \(v(u)\)  of such curves, one has to find the real root of the cubic polynomial
(setting for short \(x\equiv\mathrm{dn}(v,k')\))
\begin{eqnarray}
 P_3(x;u)&=&\kappa\,\mathrm{cd}(u)\left[k^2\bar w^2-\kappa^2\mathrm{ns}^2(u)\right]x^3
 -\bar w\,k\left[\kappa^2\mathrm{cs}^2(u)-\nu^2k'^2\right]x^2+\nonumber\\
& & +\kappa k^2\mathrm{cd}(u)\left[\bar w^2\mathrm{ds}^2(u)+\nu^2k'^2\right]x+\bar w\,k^3
\left[\bar w^2\mathrm{ns}^2(u)-\kappa^2\right].
\end{eqnarray}
Since all the coefficients are elliptic functions with periods $4 \KK$ and $2 i \KK$, the roots \(x_i(u)\), $i=1,2,3$ will also be  elliptic functions of \(u\), with the same periods, such that the polynomial itself \(P_3(x(u);u)\) will be an elliptic function. Imposing $ P_3(x;u)=0$  for any value of \(u\), implies that the poles of \(x(u)\) have to be canceled by the zeros of the 
coefficient functions of \(P_3(x;u)\). 
By studying the locus of points where the coefficient functions of $P_3(x(u),u)$ vanish, it allows us to compute
\begin{equation}
 x(u)=\frac{k \,\bar w}{\kappa}\mathrm{dc}(u,k)\,, \qquad\qquad 
 v(u)=\mathrm{dn}^{-1}\left[\frac{k\, \bar w}{\kappa}\mathrm{dc}(u,k),k'\right].
\end{equation}
For completeness, we report other special values of $\bar\Omega^2$ which appear in Fig. \ref{fig:FSPlane1}, 
\begin{eqnarray}
&& \bar\Omega^2\left(\mathbb{K}-\mathrm{sn}^{-1}\left(\frac{\bar w}{\kappa}\sqrt{\frac{\kappa^2-\nu^2}{\bar w^2-\nu^2}},k\right)\right)=
 -\left(\frac{\kappa^2}{\nu^2}-1\right),\quad
\quad
 \bar\Omega^2(\mathbb{K})=-\frac{(\kappa^2-\nu^2)(\bar w^2-\kappa^2)}{\kappa^2(\bar w^2-\nu^2)},
  \nonumber\\\nn
&& \bar\Omega^2(2\mathbb{K}+i\mathbb{K}')=\frac{(\bar w-\kappa)^2}{\bar w^2-\nu^2},\qquad
 \bar\Omega^2(i\mathbb{K}')=\frac{(\bar w+\kappa)^2}{\bar w^2-\nu^2},
\qquad \bar\Omega^2(0)=0, \qquad
 \bar\Omega^2(\mathbb{K}+i\mathbb{K}')=1-\frac{\kappa^2}{\bar w^2}\,.
\end{eqnarray}

We are now ready to illustrate the various branches for $\bar\Omega^2$. For convenience,  we define
\begin{equation}
 \chi_{\pm}(\bar\Omega)=\left(\kappa\, \bar w\pm\sqrt{(\bar w^2-\nu^2)(\kappa^2-\nu^2+\nu^2\bar\Omega^2)}\right)^2-\nu^4.
\end{equation}
For a given real value of \(\bar\Omega^2\) the linear independent solutions of the fourth order differential equation (\ref{eq:opnunonzero2}) are 
\begin{eqnarray}\label{eq:foursolutionsfolded2}
 f_{1,2}(x,\bar\Omega)&=&\frac{H(x\pm\alpha_1)}{\Theta(x)}e^{\mp x\left[Z(\alpha_1)+\lambda(\alpha_1)\right]},\\
\label{eq:foursolutionsfolded2bis}
 f_{3,4}(x,\bar\Omega)&=&\frac{H(x\pm\alpha_2)}{\Theta(x)}e^{\mp x\left[Z(\alpha_2)+\lambda(\alpha_2)\right]},
\end{eqnarray}
where the \(\alpha_i\)'s as functions of \(\bar\Omega\) have to be chosen (see Fig. \ref{fig:FSPlane1})
\begin{itemize}
\item for \(-\infty<\bar\Omega^2<-(\frac{\kappa^2}{\nu^2}-1)\) as
\begin{eqnarray}\label{alphafolded1}
\alpha_{1,2}(\Omega)&=&\mathrm{sn}^{-1}\left[\textstyle\sqrt{\frac{(\kappa^2-\nu^2)+(\bar w^2-\nu^2)(1-\bar\Omega^2)-\sqrt{((\kappa^2+\bar w^2)-(\bar w^2-\nu^2)
 \bar\Omega^2)^2-4\kappa^2\bar w^2}}{2(\kappa^2-\nu^2+\nu^2\bar\Omega^2)}},k^2\right]\pm\\
& &\pm i\,\mathrm{dn}^{-1}\left[\frac{k}{\kappa}\textstyle\sqrt{\frac{(\bar w^2-\kappa^2)^2-(\bar w^2-\nu^2)(\kappa^2+\bar w^2)\bar\Omega^2+
 (\bar w^2-\kappa^2)\sqrt{((\kappa^2+\bar w^2)-(\bar w^2-\nu^2)\bar\Omega^2)^2-4\kappa^2\bar w^2}}{-2(\bar w^2-\nu^2)\bar\Omega^2}},k'\right],
\nonumber
\end{eqnarray}
\item for \(-(\frac{\kappa^2}{\nu^2}-1)<\bar\Omega^2<-\frac{(\kappa^2-\nu^2)(\bar w^2-\kappa^2)}{(\bar w^2-\nu^2)\kappa^2}\) as
\begin{equation}\label{alphafolded2}
 \alpha_1(\bar\Omega)=\mathbb{K}-\mathrm{sn}^{-1}\left[\textstyle\sqrt{\frac{\chi_-(\bar\Omega)+\nu^4\bar\Omega^2/k^2}{\chi_-(\bar\Omega)+\nu^4\bar\Omega^2}},k\right],
\end{equation}
\item for \(-\frac{(\kappa^2-\nu^2)(\bar w^2-\kappa^2)}{(\bar w^2-\nu^2)\kappa^2}<\bar\Omega^2<0\) as
\begin{equation}\label{alphafolded3}
  \alpha_1(\bar\Omega)
 =2\mathbb{K}-\mathrm{sn}^{-1}\left[\textstyle\sqrt{\frac{-\nu^4\bar\Omega^2}{k^2\chi_-(\bar\Omega)}},k\right],
\end{equation}
\item for \(0<\bar\Omega^2<\frac{(\bar w-\kappa)^2}{\bar w^2-\nu^2}\) as
\begin{equation}
 \alpha_1(\bar\Omega)=2\mathbb{K}+i\,\mathrm{sn}^{-1}\left[\textstyle\sqrt{\frac{\nu^4\bar\Omega^2}{k^2\chi_-(\bar\Omega)+\nu^4\bar\Omega^2}},k'\right],
\end{equation}
\item for \(\frac{(\bar w-\kappa)^2}{\bar w^2-\nu^2}<\bar\Omega^2<1-\frac{\kappa^2}{\bar w^2}\) as
\begin{equation}
 \alpha_1(\bar\Omega)=2\mathbb{K}+i\mathbb{K}'-\mathrm{sn}^{-1}\left[\textstyle\sqrt{\frac{\chi_-(\bar\Omega)}{-\nu^4\bar\Omega^2}},k\right],
\end{equation}
\item for \(1-\frac{\kappa^2}{\bar w^2}<\bar\Omega^2<\frac{(\bar w+\kappa)^2}{\bar w^2-\nu^2}\) as
\begin{equation}
 \alpha_1(\bar\Omega)=\mathbb{K}+i\mathbb{K}'-\mathrm{sn}^{-1}\left[\textstyle\sqrt{\frac{\chi_-(\bar\Omega)+\nu^4\bar\Omega^2}
 {k^2\chi_-(\bar\Omega)+\nu^4\bar\Omega^2}},k\right],
\end{equation}
\item for \(\frac{(\bar w+\kappa)^2}{\bar w^2-\nu^2}<\bar\Omega^2<\infty\) as
\begin{equation}\label{C32}
 \alpha_1(\bar\Omega)=i\,\mathrm{sn}^{-1}\left[\textstyle\sqrt{\frac{\nu^4\bar\Omega^2}{k^2\chi_-(\bar\Omega)+\nu^4\bar\Omega^2}},k'\right],
\end{equation}
\item for \(-(\frac{\kappa^2}{\nu^2}-1)<\bar\Omega^2<0\) as
\begin{equation}\label{C33}
  \alpha_2(\bar\Omega)=\mathrm{sn}^{-1}\left[\sqrt{-\frac{\nu^4\bar\Omega^2}{k^2\chi_+(\bar\Omega)}},k\right],
\end{equation}
%
%
\item for \(0<\bar\Omega^2<\infty\) one has
\begin{equation}\label{alphafoldedfin}
 \alpha_2(\bar\Omega)=i\,\mathrm{sn}^{-1}\left[\textstyle\sqrt{\frac{\nu^4\bar\Omega^2}{k^2\chi_+(\bar\Omega)+\nu^4\bar\Omega^2}},k'\right].
\end{equation}
\end{itemize}
 
 In the main body we have used the following identity, which does not seem to be 
 tabulated but can be easily checked to be true
 \begin{equation}\label{Landen}
 Z(\alpha,k)=\frac{2}{1+\tilde k'}Z\left(\frac{\alpha}{1+\tilde k'}+\frac{i\mathbb{K}'}{1+\tilde k'},\tilde k\right)-
 \frac{1+\mathrm{cn}(\alpha,k)\mathrm{dn}(\alpha,k)}{\mathrm{sn}(\alpha,k)}+\frac{i\pi}{2\mathbb{K}},
\end{equation}
where $\tilde k$ is the Landen transformed modulus, {\it i.e.} $\tilde k^2= 4k/(1+k)^2$\,.

\subsection{The $\nu= 0$ limit}
\label{app:nu0}

Here we consider \eqref{alphafolded1}-\eqref{alphafoldedfin}  and  outline explicitly the   \(\nu\to 0\) limit of those $\alpha$'s  in the negative-frequency
range useful to obtain the determinant \eqref{detonu0} as a $\nu\to0$ limit of \eqref{quasimomnu}-\eqref{detbosfolded}. 
We keep in mind that
\(k^2=(\kappa^2-\nu^2)/(\omega^2-\nu^2)\).
 For the first quasi-momentum, we notice that the ellipse segments parameterized by \eqref{alphafolded1}  shrink  for \(\nu\to 0\) 
 to a point and \eqref{alphafolded1} becomes irrelevant.
  The interval corresponding to \eqref{alphafolded2} extends to \(-\infty<\bar\Omega^2<-k'^2\) and \eqref{alphafolded2}  becomes
\begin{equation}
 \alpha_1(\bar\Omega)\to \mathbb{K}-\mathrm{sn}^{-1}\left[\sqrt{\frac{(k'^2+\bar\Omega^2)^2}{(k'^2-\bar\Omega^2)^2}},k
 \right].
\end{equation}
 The interval corresponding to  \eqref{alphafolded3}  extends to \(-k'^2<\bar\Omega^2<0\) and  \eqref{alphafolded3}  becomes
\begin{equation}
 \alpha_1(\bar\Omega)\to 
 2\mathbb{K}-\mathrm{sn}^{-1}\left[\sqrt{-\frac{4\bar\Omega^2}{(1+k^2-\bar\Omega^2)^2-4k^2}},k\right]~.
 \end{equation}
 Considering the second quasi-momentum, the interval corresponding to  
 \eqref{C33}
 extends for \(\nu\to 0\) to \(-\infty<\bar\Omega^2<0\) and one gets
 \begin{equation}
 \mathrm{sn}^2(\alpha_2(\bar\Omega))\sim 
 \frac{\bar\Omega^2}{4\kappa^4}\,\nu^4~,\quad
 \mathrm{cn}^2(\alpha_2(\bar\Omega))\sim 
 1-\frac{\bar\Omega^2}{4\kappa^4}\,\nu^4~,\quad
 \mathrm{dn}^2(\alpha_2(\bar\Omega))\sim 
 1-\frac{\bar\Omega^2}{4\kappa^2\omega^2}\,\nu^4~,
\end{equation}
  so that
  \begin{equation}
    p_2\to i\,\bar\Omega\,.
    \end{equation}

\bibliographystyle{nb}
\bibliography{Ref_LL_Exact}

\end{document}